\documentclass[aps,prr,reprint,twocolumn,superscriptaddress,floatfix,nofootinbib,amsmath, amssymb]{revtex4-1}
\usepackage[utf8]{inputenc}
\usepackage{graphicx}
\usepackage{dcolumn}
\usepackage{bm}
\usepackage{color}
\usepackage[normalem]{ulem}

\def\spose#1{\hbox to 0pt{#1\hss}}
\def\lesssim{\mathrel{\spose{\lower 3pt\hbox{$\mathchar"218$}}
 \raise 2.0pt\hbox{$\mathchar"13C$}}}
\def\gtrsim{\mathrel{\spose{\lower 3pt\hbox{$\mathchar"218$}}
 \raise 2.0pt\hbox{$\mathchar"13E$}}}

\begin{document}
\title{Computing Canonical Averages with Quantum and Classical Optimizers: Thermodynamic Reweighting for QUBO Models of Physical Systems}
\author{Francesco Slongo}
\affiliation{Scuola Internazionale Superiore di Studi Avanzati (SISSA), Via Bonomea 265, I-34136 Trieste, Italy.}
\author{Cristian Micheletti}
\email{cristian.micheletti@sissa.it}
\affiliation{Scuola Internazionale Superiore di Studi Avanzati (SISSA), Via Bonomea 265, I-34136 Trieste, Italy.}

\begin{abstract}
{\bf Abstract.}
We present a general method to compute canonical averages for physical models sampled via quantum or classical quadratic unconstrained binary optimization (QUBO). First, we introduce a histogram reweighting scheme applicable to QUBO-based sampling constrained to specific intervals of an order parameter, e.g., physical energy. Next, we demonstrate that the scheme can accurately recover the density of states, which in turn allows for calculating expectation values in the conjugate ensemble, e.g., at a fixed temperature. The method can thus be used to advance the state-of-the-art characterization of physical systems that admit a QUBO-based representation and that are otherwise intractable with real-space sampling methods. A case in point are space-filling melts of lattice ring polymers, recently mapped in QUBO form, for which our method reveals that the ring catenation probability is non-monotonic with the bending rigidity.
 \end{abstract}
\date{\today}
\maketitle

\noindent{\bf \large Introduction}
The possibility of evaluating multiple solutions concurrently makes quantum computing ideally poised to solve optimization tasks in ways that are radically different from conventional methods. A prominent class of combinatorial problems solvable with quantum algorithms is quadratic unconstrained binary optimization (QUBO), which includes SAT, maximum clique, and graph coloring.

Solving a QUBO problem is equivalent to finding the ground state configuration(s) of a quadratic Ising-like Hamiltonian:
\begin{equation}
    \label{eqn:QUBO_Ising}
    \mathcal{H}_0(\sigma) =  \sum_i h_i\, \sigma_i\ + \sum_{i\neq j} J_{ij}\, \sigma_i \sigma_j\ ,
\end{equation}
\noindent where the $h$ variables are the local external fields, $J$ is a symmetric interaction matrix, and the $\sigma$s are binary variables taking values of either 0 or 1.

In pursuit of potential advantages offered by quantum optimization and simulations \cite{zhang2017advantages,albash2018demonstration,kumar2020achieving,mazzola2021sampling,layden2023quantum,abbas2023quantum,sandt2023efficient,mazzola2024quantum,mcardle2020quantum,daley2022practical}, and to leverage the practical speedups of special-purpose classical Ising machines \cite{belletti2008janus,belletti2008nonequilibrium,matsubara2018ising,mohseni2022ising}, researchers are increasingly recasting conventional sampling simulations of discrete physical systems as QUBO problems \cite{perdomo2012finding,lucas2014ising,hernandez2017enhancing,xia2017electronic,li2018quantum,harris2018phase,king2018observation,mulligan2019designing,streif2019solving,hauke2020perspectives,willsch2020support,terry2020quantum,kitai2020designing,carnevali2020vacancies,micheletti2021polymer,hatakeyama2021tackling,yarkoni2022quantum,irback2022folding,slongo2023quantum,baiardi2023quantum,irback2024using}. The mapping typically involves a one-to-one correspondence of the real-space configurations of the original system and the degenerate ground state solutions of an appropriate QUBO Hamiltonian. Uncorrelated configurations of the physical systems can thus be obtained by performing independent and unbiased minimizations of the QUBO Hamiltonian in the abstract space of its binary variables, followed by backmapping the solutions to real-space representations.

For soft matter systems, such as dense phases of polymers, resorting to the QUBO-based sampling on fully-quantum or hybrid classical-quantum annealers can be significantly more efficient than conventional Monte Carlo sampling in real space \cite{slongo2023quantum}. Remarkably, the speedup benefits can be reaped, albeit to a lesser degree, even when the QUBO minimization is performed classically. A case in point is sampling maximally dense lattice ring polymers with infinite bending rigidity.
Compared to real-space advanced sampling methods, numerical benchmarks show that classical QUBO-based sampling takes the runtime scaling with system size, $N$, from $N^{5.4}$ down to $N^{3.7}$\cite{slongo2023quantum}.

QUBO-based sampling thus holds the promise of bringing significant performance improvements compared to Monte Carlo methods to many other instances of hard physical problems \cite{perdomo2012finding,lu2019quantum,robert2021resource,irback2022folding,ghamari2022sampling,linn2024resource}, including protein design strategies and RNA secondary structure prediction \cite{fox2022rna,irback2024using,panizza2024protein}.
However, one known limitation is that QUBO-based sampling is natively tied to the microcanonical ensemble, unlike Monte Carlo methods, which naturally operate in the canonical ensemble. This is because QUBO schemes, as long as they are free of biases \cite{moreno2003finding,mandra2017exponentially}, allow for covering the ground state manifold uniformly so that different energy-minimizing states are sampled with equal probability.

This limitation can be partly overcome in specific contexts, e.g., when excited states returned by quantum annealing runs are informative about the low-temperature Gibbs ensemble of the physical model \cite{vuffray2022programmable,nelson2022high,ghamari2022sampling,sandt2023efficient,layden2023quantum}. However, even aside from considerations of fair sampling\cite{mandra2017exponentially,konz2019uncertain,pochart2022challenges}, a general approach for calculating Boltzmann averages for QUBO representations of physical systems is still lacking.

In response to this challenge, here we introduce a QUBO-based scheme that enables computing canonical expectation values by combining the use of slack variables and a thermodynamic reweighting scheme.
First, slack variables are introduced in the Hamiltonian to restrain the order parameter of interest, e.g., the energy of the physical model, within specific intervals. Next, a suitable weighted histogram method is used to combine data from overlapping intervals and thus recover the density of states.
The latter is finally used to calculate expectation values in the conjugate ensemble, e.g., at a fixed temperature.

The approach is not restricted to computing energy-like expectation values and is, in principle, applicable to computing canonical averages of any order parameter. Accordingly, the method can be advantageously used on physical systems that are more tractable in QUBO form than in the native representation. For such systems, combining the intervalled QUBO-based sampling and reweighting scheme can significantly advance the characterization of the canonical equilibrium properties beyond state-of-the-art sampling methods such as real-space Monte Carlo.

We demonstrate this by considering space-filling melts of ring polymers. The system is, at one time, of broad interest in soft matter physics as well as a paradigm of the challenges of real-space sampling due to rapidly increasing autocorrelation times with system size and density. We show that our approach enables the first systematic characterization of inter-molecular linking of the rings as a function of their bending rigidity, revealing a non-monotonic relationship.

The findings highlight the method's potential for providing breakthrough insights for QUBO-based physical models. They also motivate expanding the range of physical models mapped in QUBO form, where one could further harness the rapid development of optimization platforms using quantum algorithms and hardware.

\section{Multi-histogram reweighting for QUBO}
\subsection{Targeting QUBO-based sampling at energy intervals}
We consider a QUBO-based encoding of a discrete physical system, represented by a Hamiltonian $\mathcal{H}$ whose ground states are in one-to-one correspondence with the admissible configurations of the physical system. As in eq.~\ref{eqn:QUBO_Ising}, $\mathcal{H}$ can include up to quadratic interactions of the variables $\sigma$, which take values 0 or 1. Throughout this manuscript, we interchangeably refer to the $\sigma$s as binary or spin variables, although in quantum computing contexts the latter term is commonly reserved for variables taking values $\pm 1$.

We further assume that the energy $E$ of the physical system -- distinct from the QUBO Hamiltonian $\mathcal{H}$ -- can be written as a linear combination of the spin variables with integer coefficients:
\begin{equation}
    E = E_0+ \Delta E \sum_i a_i \sigma_i
    \label{eqn:spectrum}
\end{equation}
where the $a_i$s are arbitrary integers. Writing the physical energy in the form of eq.~\ref{eqn:spectrum} may require introducing suitable ancilla spin variables \cite{leib2016transmon}.
Without loss of generality, in the following, we will set $E_0=0$ and $\Delta E=1$ unless otherwise stated. Specific examples for Ising and lattice polymer models will be discussed in Sections \ref{sec:validation} and \ref{sec:application}.

Plain QUBO-based sampling can be directed and restricted to single values of the discrete physical energy by complementing $\mathcal{H}_0$ of eq.~\ref{eqn:QUBO_Ising} with a quadratic constraint that penalizes deviations of $E$ from $\bar{E}$:
\begin{equation}
    \mathcal{H} \left( \sigma \right) = \mathcal{H}_0(\sigma) + A \left( \sum_i a_i \sigma_i - \bar{E}\right)^2\ ,
    \label{eqn:singleenergy}
\end{equation}
with $A>0$.
Because the added constraint is quadratic, $\mathcal{H}$ is still a QUBO Hamiltonian. Its ground states manifold now provides a one-to-one coverage of the microcanonical ensemble at energy $\bar{E}$.

In principle, this energy targeting could be repeated for all admissible values of $\bar{E}$. However, even doing so would not enable the calculation of canonical averages since the necessary knowledge of the entropic weight of states at different energies can be gained only by targeting $E$ intervals spanning multiple energy levels rather than just a single value of $\bar{E}$.

To direct the QUBO-based sampling to a specific energy interval, we extend the space of spin variables to include a set of so-called slack variables, $s_0,\ s_1, ...,\ s_{m-1}$. These variables are incorporated into a quadratic constraint that generalizes the one of eq.~\ref{eqn:singleenergy}:
\begin{equation}
    \mathcal{H} \left( \sigma \right) = \mathcal{H}_0(\sigma) +  \left( \sum_i a_i \sigma_i - \bar{E} - \sum_{k=0}^{m-1} 2^k s_k \right)^2 \ .
    \label{eqn:energyinterval}
\end{equation}
\noindent Minimizing the above Hamiltonian enables the uniform sampling of microstates across the energy interval $ E_{\rm min} \le E \le E_{\rm max}$, where $E_{\rm min}=\bar{E}$ and $E_{\rm max} =\bar{E} + \left(2^{m} - 1\right)$.
Systematically varying $\bar{E}$ allows for covering the energy spectrum of the physical system in its entirety.

\subsection{Density of states from energy-intervalled sampling}
The energy histograms obtained by sampling multiple overlapping energy intervals contain, in principle, sufficient information to reconstruct the density of states, $W(E)$, of the physical system. However, $W(E)$ cannot be reconstructed with standard weighted histogram analysis methods \cite{ferrenberg1989optimized,swendsen1993modern,tuckerman2023statistical}. This is because weighted histogram methods are conceived for samples drawn from numerous Boltzmann or Boltzmann-like ensembles, a condition fundamentally different from our approach. In our framework, minimizing eq.~\ref{eqn:energyinterval}  yields samples that lack Boltzmann-like statistical weights and, especially, are restricted to preassigned energy intervals of given widths and boundaries.

We use variational principles to derive a generalized weighted histogram analysis framework to recover $W(E)$ in this new context. This new method uses data sampled from multiple staggered energy intervals to optimally reconstruct $W(E)$ with gapless coverage of the energy range of interest.

We will use the following notation. The index $j$ labels the energy intervals, defined by $E_{\rm min}(j) \leq E \leq E_{\rm max}(j)$. The total number of microstates sampled for the $j$th interval is indicated by $N_j= \sum_E n_j(E)$, where $n_j(E)$ is the number of microstates with energy $E$ sampled for interval $j$.

The above quantities can be related to the density of states via:
\begin{equation}
 \frac{\left\langle n_j(E) \right\rangle}{N_j}  =
 \begin{cases}
    \frac{W(E)}{Z_j}, & \text{if $E_{\rm min}(j) \le E \le E_{\rm max}(j)$}.\\
    0, & \text{otherwise}.
  \end{cases}
  \label{eqn:n_and_W}
\end{equation}
\noindent where the $\langle \rangle$ brackets denote the average over various realizations (sampling runs) and:
\begin{equation}
    Z_j=\sum_{E_{\rm min}(j) \le E \le E_{\rm max}(j)} W(E)
    \label{eqn:Z_definition}
\end{equation}

Thus, by inverting the relationship of eq.~\ref{eqn:n_and_W} one has that the ratio $n_j(E)/{N_j}$ provides an estimator of $W(E)$ up to a multiplicative constant.

Note that an independent estimator for $W(E)$ can be obtained for each of the intervals covering the same energy level, $E$. These independent estimates can be combined into an optimal weighted average:
\begin{equation}
W(E) = \sum^\prime_j \alpha_j(E) \frac{n_j(E)}{N_j} \, Z_j \ ,
  \label{eqn:multi}
\end{equation}
\noindent where the prime indicates that the sum is restricted to intervals that include $E$. For our purposes, the optimal $\alpha$s are those minimizing the error on the weighted sum. The variational calculation yields:
\begin{equation}
\alpha_j \propto \frac{ 1}{ \delta n_j(E)^2 } \cdot  \frac{N_j^2}{\, Z_j^2 }\ ,
\label{eqn:weights}
\end{equation}
\noindent where $\delta n_j(E)^2$ is the variance of $n_j(E)$, and the proportionality coefficient is fixed by the constraint $\sum^\prime_j \alpha_j=1$.

Assuming that the microstates in each interval are sampled uniformly and using binomial statistics arguments, the variance of $n_j(E)$ can be written as:
\begin{equation}
    \delta n_j(E)^2 = g_j \langle n_j(E) \rangle \left(1 - \frac{\langle n_j(E) \rangle}{N_j} \right)
    \label{eqn:binomial}
\end{equation}
\noindent where the factor $g_j = 1 + 2 \tau_j$ accounts for the possible finite autocorrelation time,  $\tau_j$, of the microstate sampling process \cite{muller1973dynamic}. When samples are uncorrelated, e.g.~when obtained from independent quantum annealing runs, then $g_j =1$.

Combining expressions of eqs.~\ref{eqn:n_and_W}, \ref{eqn:weights} and \ref{eqn:binomial} yields:
\begin{equation}
    W(E) = \frac{\sum_j^\prime \frac{n_j(E)}{g_j \left(1 - \frac{W(E)}{Z_j} \right)}}{\sum_k^\prime \frac{N_k}{Z_k g_k \left(1 - \frac{W(E)}{Z_k} \right)}} \ .
  \label{eqn:final}
\end{equation}
Note that the density of states can be reconstructed only up to a multiplicative constant, which can be set by imposing the normalization condition:
\begin{equation}
    \sum_E W(E) = 1 \ .
    \label{eqn:normalization_condition}
\end{equation}

The sought $W$s in the desired energy range are obtained by treating eq.~\ref{eqn:final} as a set of self-consistent equations to be solved iteratively. The procedure, its algorithmic formulation and convergence are presented in Section 1 of the Supplementary Material.

From the reconstructed $W(E)$ profile, the canonical expectation value of a generic observable, ${\cal O}$, at temperature $T$ can be obtained via thermodynamic reweighting \cite{vernizzi2020multicanonical,tuckerman2023statistical}:
\begin{equation}
    \langle {\cal O}\rangle_T = \frac{\sum_{E} \langle {\cal O} \rangle_E \, \,W(E) e^{-\beta E}}{\sum_{E} W(E) e^{-\beta E}}
    \label{eqn:expectation}
\end{equation}
\noindent where $\beta=(k_B\, T)^{-1}$ and $\langle {\cal O} \rangle_E$ and $\langle {\cal O} \rangle_T$ are the averages of ${\cal O}$ at fixed energy $E$ and temperature $T$, respectively. A summand-sorting procedure prior to the summation is advised for numerical
precision.

We note that, though the optimally weighted averaging of eq.~\ref{eqn:weights} is inspired by conventional histogram reweighting methods \cite{ferrenberg1989optimized,swendsen1993modern,tuckerman2023statistical}, the two reconstruction approaches are fundamentally different. In the conventional case, recovering the density of states requires undoing the Boltzmann weight of overlapping histograms, collected at different temperatures, covering the energy range of interest. The position and width of the histograms cannot be enforced {\em a priori} because they are system- and temperature-dependent and hence need to be chosen through tentative or pre-conditioning runs. Instead, in our scheme, the sampling can be expressly directed at user-defined energy intervals of desired width position, thus facilitating the coverage of the energy range of interest.

\subsection{Error analysis}

To estimate the errors of the reconstructed $W(E)$, we resort to a block-like analysis that makes it straightforward to account for the correlations between the $n_j$s, $Z$s, and $W$s.
For simplicity, we assume that all addressed energy intervals are covered with the same "sampling depth," $N_j=N$, and that the sampled states are uncorrelated ($\tau_j=0)$.
For each interval, we subdivide the samples into $s$ blocks of equal size, $N/s$. Next, we carry out $s$ reconstructions of the density of states, using for each interval the first block, then the second, and so on, obtaining a total of $s$ independent (normalized) profiles, $W_1(E), W_2(E), \dots, W_s(E)$. The final $W$ profile and its statistical uncertainty are obtained by computing the mean of the $W_{i=1,\dots,s}$ and the associated statistical error for each value of $E$.

\subsection{Validation}

\label{sec:validation}

To validate the QUBO-based reconstruction of $W(E)$, we applied it to an exactly solvable physical system. To this end, we considered the Ising model on the $L \times L$ square lattice with uniform nearest-neighbour spin interactions and periodic boundary conditions. For even $L$, the density of states can be calculated exactly with the recursive enumeration method of ref.~ \cite{beale1996exact}. Because the energy of the system is solely determined by the number of parallel neighboring spins, $2n_{\shortparallel}$, we used $n_{\shortparallel}$ as the natural variable for profiling $W$, and mapped the admissible configurations on the QUBO model detailed in Appendix A.

The admissible range of $n_{\shortparallel}$ is bound by 0 and $L^2$, corresponding respectively to the antiferromagnetic and ferromagnetic states. $W(n_{\shortparallel})$ is unimodal and symmetric with respect to the midpoint $n_{\shortparallel}=L^2/2$.
$W$ is minimum at the boundaries, where the ferromagnetic and antiferromagnetic states account for two configurations each, and is maximum at the midpoint.

For a stringent validation, we considered the $12\times 12$ Ising system ($L=12$), where the  "dynamic range" of $W$ spans across more than 40 orders of magnitude, making it impossible to recover the entire profile of $W(n_{\shortparallel})$ with simple sampling schemes.
For the reconstruction, we used $m=1$, $2$, and $3$ slack variables, corresponding to $n_{\shortparallel}$ intervals of width $2$, $4$, and $8$, respectively, and sampled the ground state manifold using a parallel tempering scheme (see Section 2 of the Supplementary Material).
We leveraged the unbiassed nature of classical parallel tempering-based optimizations to decouple the validation of the reconstruction method from potential external biases. These include those that can arise from, e.g.~fair-sampling issues in quantum optimizers\cite{mandra2017exponentially,konz2019uncertain,pochart2022challenges},and out-of-equilibrium effects in classical annealers\cite{moreno2003finding}.

As illustrated in Appendix A, the reconstructed profiles are practically indistinguishable from the exact one at all $m$s, and
the observed agreement is remarkably consistent throughout the broad range spanned by $W$, despite its wide dynamic range.
Furthermore, the reported data demonstrate that the block analysis provides a reliable and unbiased estimate of the statistical uncertainty of the reconstructed $W$.

\section{Application: Adding bending rigidity to ring polymer melts}
\label{sec:application}

We now apply the approach to characterize the topological entanglement in equilibrated melts of semiflexible, topologically unrestricted ring polymers. The canonical sampling of such a system is challenging for conventional approaches based on real-space representations, such as molecular dynamics or Monte Carlo simulations. Consequently, no results are available for how knotting and linking probabilities change with the rings' bending rigidity.

These questions are relevant across diverse contexts, from biology and physical chemistry to material science and soft matter physics.
For instance, polymer ring melts are key reference systems for understanding the unique features of the multiscale genome organization \cite{grosberg1993crumpled,rosa2014ring,abdulla2023topological}, and how topoisomerase enzymes influence it \cite{piskadlo2017metaphase,roca2022keeping}.
Additionally, modern interpretations of density-induced phase transitions in liquids are related to entanglements of closed paths along the chemical bonds of the system \cite{neophytou2022topological,neophytou2024hierarchy}. Dense systems of circular molecules are also crucial for prospective realization of topological meta-materials, including Olympic gels and self-assembled molecular chainmails \cite{luengo2024shape,klotz2020equilibrium,polles2016optimal,chiarantoni2023linear,liu2022polycatenanes,gil2015catenanes,rauscher2020dynamics,rauscher2020thermodynamics}.
Finally, in soft matter contexts, melts of ring polymers are at the heart of ongoing endeavors to understand anomalous relaxation dynamics and response to shear flow \cite{ito2007novel}, aging of active topological glasses \cite{smrek2020active}, and how it is affected by both chain density and topology \cite{o2020topological,smrek2020active,farimani2024effects,micheletti2024topology}.

The computational challenges in tackling these systems arise from their high packing density\cite{schmid2022understanding}. The latter dramatically hinders the physical relaxation of the chains, making it computationally prohibitive to obtain uncorrelated equilibrated samples using molecular dynamics simulations. An analogous slowing down affects Monte Carlo simulations, too. By employing non-local (non-physical) moves, Monte Carlo methods can often speed up the sampling compared to molecular dynamics. However, the acceptance rate of such non-local moves drops significantly with increasing system density, again due to steric clashes. The problem is exacerbated when going from fully flexible to semi-rigid and finally to rigid rings. These challenges affect continuum and lattice polymer models alike.

\subsection{QUBO-based reconstruction of the density of states}

Prompted by these considerations, we addressed the entanglement of canonically equilibrated self-assembled melts of lattice ring polymers for arbitrary values of their bending rigidity $\kappa_b$.
We adopted the QUBO-based formulation of lattice polymers introduced in refs.~ \cite{micheletti2021polymer,slongo2023quantum}, which enables the sampling of self-assembled ring melts on regular and irregular lattices, and considered the stringent case of ring melts that fill cubic lattices of $N$ sites.

As detailed in Appendix B, the QUBO Hamiltonian for such systems can be formulated in terms of two types of Ising-like binary variables attached respectively to individual lattice edges and pairs of incident edges meeting at $\pi/2$ angles. The two sets of variables are introduced to keep track of (i) the lattice edges that are occupied by the bonds of the self-assembled rings and (ii) the number of corner turns of the rings, $n_c$.

Note that on hypercubic lattices, as in the considered case, $n_c$ is proportional to the total curvature of the ring melt up to a $\pi/2$ factor. It is thus the parameter of choice for the density of states, $W(n_c)$, required to compute the desired canonical expectation values as a function of the conjugate variable, i.e.~the ring's bending rigidity, $\kappa_b$.

\begin{figure*}[ht!]
    \centering
    \includegraphics[width=0.8\textwidth]{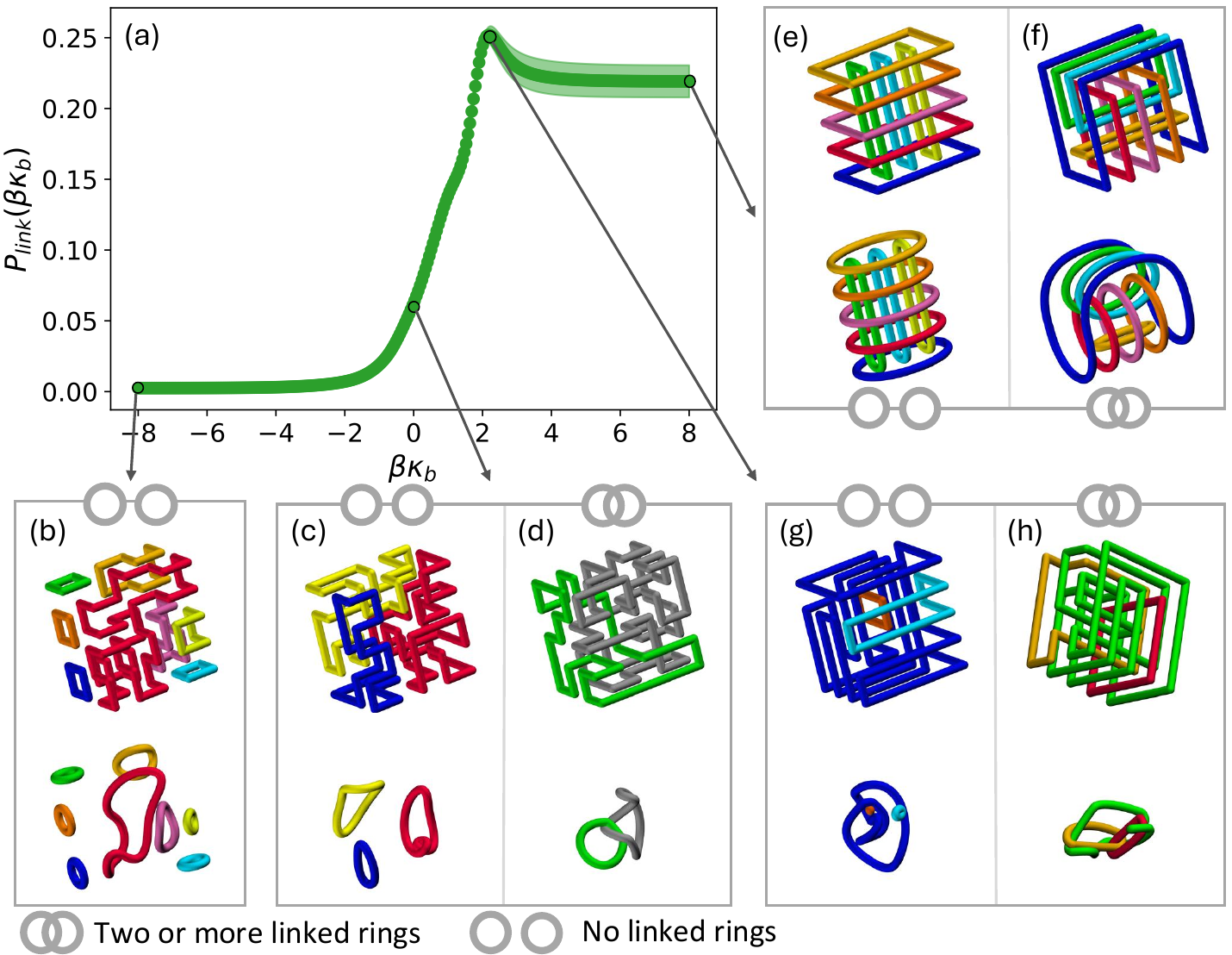}
    \caption{{\bf Equilibrium linking probability of melts of self-assembled rings as a function of the reduced bending rigidity $\beta \kappa_b$.}   (a) Linking probability $P_{\rm link}$ for a canonically-equilibrated ring melt filling a $5 \times 5 \times 4$ cubic lattice. The data points represent the average $P_{\rm link}$ computed from $4$ reconstructions and the shaded bands indicate the standard error of the mean. The considered range of $\beta \kappa_b$ spans from -8 to +8, i.e.~from strongly favoured to strongly suppressed corner turns. Across this range, the linking probability is non-monotonic, peaking at $\beta \kappa_b \sim 2$. Panels (b-h) show typical configurations at the values of $\beta \kappa_b$ indicated by the callouts. The configurations were randomly picked among those with $n_c$ equal to the average curvature at the corresponding value $\beta \kappa_b$.}
    \label{fig:linkingprobability}
\end{figure*}

We reconstructed $W(n_c)$ for space-filling ring melts on a $5 \times 5 \times 4$ cubic lattice for a total of 100 lattice sites; see (b-h) panels in Fig.~\ref{fig:linkingprobability}.
To cover the entire admissible range $32 \leq n_c \leq 100$, we used multiple intervalled samplings with unit increment in $\bar{n}_c$ and $m = 3$ slack variables, corresponding to intervals of widths 8. Considering the numerous intervals involved, we performed the sampling with a parallel tempering scheme on a classical high-performance computing cluster, and reserved the use of quantum optimization platforms to the more manageable $3\times3\times2$ system, which we discuss later.

\begin{figure}[!ht]
    \centering
    \includegraphics[width=0.45\textwidth]{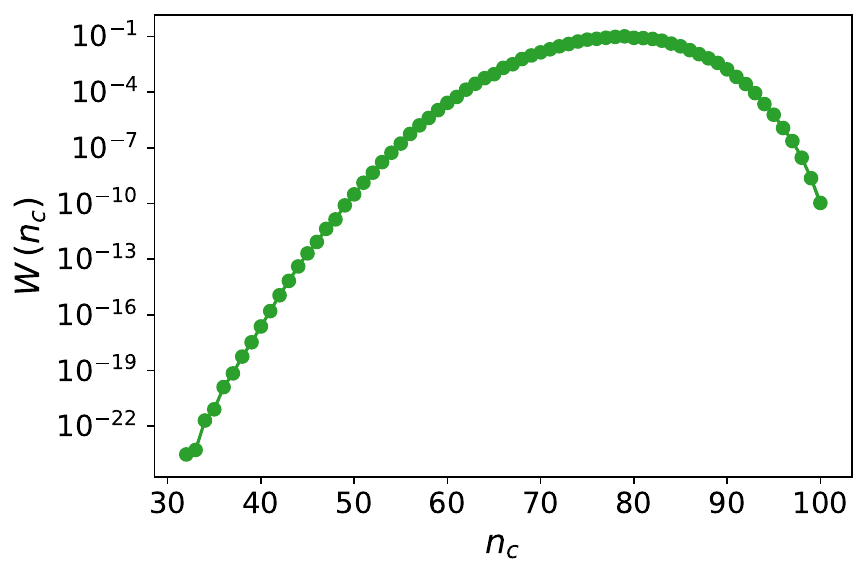}
    \caption{{\bf Density of states as a function of total curvature for a ring melt filling a $5\times5\times4$ lattice.} The reconstruction was obtained using a sampling depth $d \approx 1,000$ for each of the $76$ intervals of length $8$ ($m=3$) straddling the entire range of curvature values, $ 32 \le n_c \le 100$, with the first interval starting at 26 and the last one ending at 106. The data points represent the average density of states computed from $4$ reconstructions; error bars are smaller than the symbol size, the average relative error being around $5\%$.}
    \label{fig:densitypolymer}
\end{figure}

The density of states obtained using $m=3$ variables is shown in Fig.\ref{fig:densitypolymer}.

$W\left(n_c\right)$ is peaked around $n_c=78$, which would be the most probable value of $n_c$ for curvature-unrestricted sampling. $W$ decreases rapidly on both sides of the peak, dropping by more than 9 and 22 orders of magnitude to the right and left, respectively.

We note that if the same total number of states were collected from a curvature-unrestricted QUBO sampling, rather than from multiple overlapping intervals, then the single resulting histogram would only cover the $60 \leq n_c \leq 96$ range. This limited coverage would preclude the calculation of the entire $W$ profile and, hence, of canonical expectation values at arbitrary $\beta \kappa_b$.

\subsection{Canonical expectation values for varying bending rigidity}\label{sec:2B}

The density of states $W(n_c)$ enables calculating the expectation values of a generic observable, $\mathcal{O}$, at fixed inverse temperature, $\beta$, and bending rigidity, $\kappa_b$, via:
\begin{equation}
    \langle \mathcal{O} \rangle_{\beta \kappa_b} = \frac{\sum_{n_c} \langle \mathcal{O} \rangle_{n_c} W\left(n_c\right) e^{- \beta \kappa_b n_c}}{\sum_{n_c} W\left(n_c\right) e^{- \beta \kappa_b n_c}}\ ,
    \label{eqn:Observable_final}
\end{equation}

\noindent where $\langle \mathcal{O} \rangle_{n_c}$ is the average of the observable of interest computed at fixed $n_c$.
In the above expression, which specializes eq.~\ref{eqn:expectation} to self-assembled ring melts, we have set $E = \kappa_b n_c$, thereby absorbing the curvature of $\pi/2$ at each corner turn into the definition of $\kappa_b$.

Note that $W$ is the same for any considered observable and, therefore, needs to be reconstructed only once by covering the admissible range of $n_c$ with overlapping intervals. In addition, it is advantageous to compute the constrained average, $\langle \mathcal{O} \rangle_{n_c}$, using (uncorrelated) states with the given $n_c$ from all intervals that include $n_c$.

\subsection{Results: linking probability}
We used the above scheme to profile the linking (concatenation) probability of canonically equilibrated melts of rings as a function of the bending rigidity.

Concatenation constraints, also termed mechanical bonds, are key structural motifs for supramolecular self-assemblies, from self-limited synthetic catenanes \cite{chichak2004molecular,beves2011strategies,polles2016optimal,marenda2018discovering} to modular ones \cite{niu2009polycatenanes,wu2017poly,datta2020self,chiarantoni2022effect,chiarantoni2023linear}, including three-dimensional Olympic gels \cite{speed2024assembling}.

In recent years, various experimental advancements \cite{yamamoto2015cyclic,yamamoto2016light}, especially in metal-ion templating techniques, have finally made it possible not only to externally control the geometry and topology of mechanical bonding but also to boost the yield of these topological constructs. However, despite these breakthroughs, identifying the conditions most conducive to inter-molecular linking remains an open problem.

In this regard, the bending rigidity, $\kappa_b$ is a natural parameter for the design of topologically-bonded materials.
However, studying the effect of $\kappa_b$ on concatenation probability of long and densely packed rings is challenging for real-space Monte Carlo and molecular dynamics because the autocorrelation times increase rapidly with the rings' rigidity. For instance, high effective bending rigidities, e.g., due to electrostatic repulsions, can cause the dynamical arrest in ring melts \cite{staňo2023cluster}; in addition, concatenation constraints significantly slow down the system's relaxation dynamics \cite{datta2024influence}. Consequently, profiling inter-chain linking in canonical equilibrium has so far been feasible only for a few distinct values of $\beta \kappa_b$ and for discrete models amenable to special-purpose sampling methods \cite{ubertini2023topological}.
Thus, how mutual entanglements in ring melts vary as a function of $\beta \kappa_b$ remains an unsolved problem for conventional Monte Carlo methods. For the same reasons, it is the natural avenue for applying the QUBO-based sampling and the thermodynamic reweighting technique. In fact,  our recent study of ref.~\cite{slongo2023quantum} has shown the potential of using QUBO-based models to profile the entanglement of ring melts. However, such considerations were  restricted to the microcanonical constant-curvature ensemble, preventing to draw any conclusion for the conjugate canonical ones as a function of $\beta \kappa_b$.

Accordingly, we computed the $\beta\kappa_b$ dependence of the linking probability, $P_{\rm link}$, defined as the probability that self-assembled states contain at least one linked pair of rings. To this end, we used eq.~\ref{eqn:Observable_final} after identifying $\langle \mathcal{O} \rangle_{n_c}$ in with $P_{\rm link}(n_c)$.

The resulting $P_{\rm link}(\beta \kappa_b)$ is shown in Fig.~\ref{fig:linkingprobability}a. The shaded band denotes the estimated statistical error on the average.
Notice that $\beta \kappa_b$ is varied continuously and that positive and negative values of the bending rigidity can be seamlessly considered. The negative bending rigidity case, corresponding to situations where bending is energetically favored, is addressed rarely in conventional simulations.

The data in Fig.~\ref{fig:linkingprobability}b establishes a remarkable novel result, namely that the linking probability has a unimodal dependence on $\beta \kappa_b$.

This nonmonotonicity is best discussed considering the limiting cases where $\beta\,\kappa_b$ takes on large negative and positive values.

For $\beta \kappa_b < -4$, $n_c$ is maximum, corresponding to a $\pi/2$ turn at each lattice site. The resulting rings are so tightly wound that they do not leave openings that can be threaded by the other rings thereby preventing linking. In the opposite case, $\beta \kappa_b > 4$, typical configurations correspond to nested stacked rings -- similar to the columnar structures observed in concentrated solutions of semiflexible ring polymers \cite{poier2016anisotropic} --  as illustrated in Fig.~\ref{fig:linkingprobability}e. The rings in these columnar structures are planar and, hence, free of intra-chain entanglement (knotting). However, their inter-chain entanglement (linking) remains possible in the form of interlocked ring stacks, as shown in Fig.~\ref{fig:linkingprobability}f. Consequently,  $P_{\rm link}$ attains a finite value for $\beta \kappa_b \gg 0$.
Remarkably, the two limits are bridged non-monotonically, with the maximum linking probability occurring at $\beta \kappa_b \sim 2$.

We conclude that a finite bending rigidity is required to balance two opposite effects: on the one hand, rings must not be too flexible or meandering because some degree of directional persistence is necessary to form loops wide enough to be threadable. On the other hand, while a large stiffness does produce configurations that could be threadable, it also suppresses configurational entropy, and hence it is not optimally conducive to linking.

The generality and robustness of the nonmonotonicity of $P_{\rm link}(\beta \kappa_b)$ is indicated by the fact that it does not depend on whether the number of self-assembled rings is allowed to fluctuate or is set equal to 3 or more rings by {\em a posteriori} selection, see Appendix B and Fig.~S1.

This behavior has not been previously reported or observed in polymer systems at large.
The result significantly advanced the understanding of how bending rigidity influences the topological entanglement of polymer systems.
Previous studies have been limited to single, isolated polymers, where the only possible form of topological entanglement is intra-chain (i.e., knotting), which is also unimodal with respect to bending rigidity \cite{coronel2017non}. Interestingly, we found an analogous knotting property for rings that are not isolated but are part of self-assembled melts, as shown in Appendix B. The shared unimodality of both linking and knotting probabilities reveals a previously unrecognized connection and common microscopic basis between the intra- and inter-ring entanglements in spite of their otherwise very distinct nature.

We expect that the unimodality of $P_{\rm link}$, here established for a maximally dense system of ring melts, ought to manifest more broadly, e.g., at partial space-filling and in various realizations of supramolecular self-assemblies or topologically-unrestricted ring polymers. The applicative potential of the result to boost the inherently low yield of molecular interlockings in supramolecular self-assemblies \cite{gil2015catenanes,polles2016optimal}. Our results indicate that a judicious design of the bending rigidity of the circular elements could afford considerable latitude for tuning and maximizing the concatenation probability.
For instance, in the case of Olympic gels assembled from individually circularizable linear DNAs \cite{speed2024assembling}, two such control parameters would be the ionic strength/valency of the solution and the DNA length, which can influence the effective rigidity by modulating the DNA persistence length and the number of Kuhn lengths, respectively.

\section{Application to Quantum Annealers}
Compared to conventional sampling methods, such as Monte Carlo of molecular dynamics with real-space polymer representations, the QUBO formulation can offer significant speedups with both classical and quantum optimizers. This advantage was demonstrated in ref. \cite{slongo2023quantum}, which examined the computational cost of sampling space-filling ring melts with minimum curvature as a function of the system size (total ring length), $N$. For ad hoc optimized real-space sampling, the computational cost scaled as $N^{5.4}$, while QUBO-based sampling improved the scaling to $N^{3.7}$ with classical annealers and to $N^{3.2}$ with hybrid classical-quantum ones\cite{slongo2023quantum}.

\begin{figure}[ht!]
    \centering
    \includegraphics[width=1\linewidth]{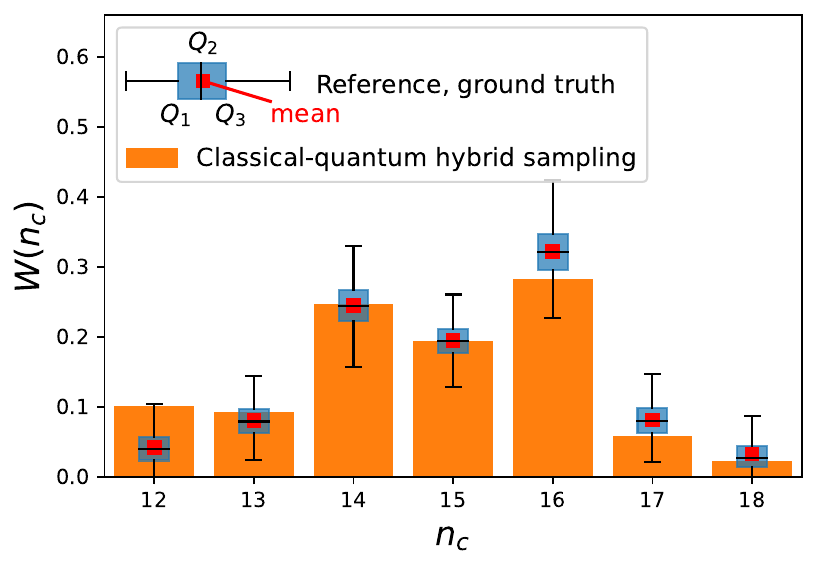}
    \caption{{\bf Validation of the density of states reconstructed from classical-quantum hybrid sampling of a ring melt filling a $3 \times 3 \times 2$ lattice.} The reconstruction (orange histograms) was computed from samples obtained with the D-wave hybrid sampler, using a sampling depth $d=50$ for each of the $4$ intervals of length $4$ ($m=2$) spanning the entire range of curvature values, $ 12 \le n_c \le 18$, with the first interval starting at $10$ and the last one ending at $20$. The ground truth normalized density of states was obtained by exhaustive enumeration of all possible states. The box plots represent the probability distribution of $500$ independent reconstructions obtained via unbiased sampling of the exhaustively enumerated set, using the same intervals and depth $d$ as with the hybrid sampler. $Q1$, $Q2$ and $Q3$ mark the first, second (median), and third quartiles, respectively. The whiskers extend to the furthest data points within 1.5 IQR of the box edges, where IQR is the interquartile range ($Q_1$ to $Q_3$).}
    \label{fig:Wcomparison}
\end{figure}

The implications are twofold. On the one hand, QUBO-based sampling of soft matter can already be advantageous on classical machines, for which computing power is widely available. On the other hand, as the size and power of quantum machines continue to advance, integrating quantum annealers and QUBO-based models could become a relevant and performative tool for physical systems where real-space MC schemes are hindered by the rapid growth of autocorrelation times with system size.

Towards these prospective applications, we used the D-Wave implementation of quantum annealers to assess their potential for reconstructing the density of states without biases arising from fair-sampling issues \cite{mandra2017exponentially,konz2019uncertain,pochart2022challenges}.

For such proof of concept demonstration, we considered ring melts in a $3 \times 3 \times 2$ cuboid.
This system size was chosen because it is sufficiently large to feature hundreds of distinct states across a certain curvature range, $ 12 \le n_c \le 18$. At the same time, the entire conformational ensemble can be explored with exhaustive enumeration methods, which we leveraged to establish the exact density of states and the statistical confidence intervals of its reconstructions at a given sampling depth.

For the reconstruction, we used $m=2$ slack variables to cover the above-mentioned $n_c$ range with $4$ staggered intervals of 4 bins each. We set the sampling depth of each interval to 50, thus maintaining coverage well below the exhaustive limit. The minimization of the QUBO Hamiltonians, which involved $115$ qubits, was performed with the D-Wave hybrid classical-quantum sampler with the default runtime of 3s. Although this was the minimum runtime allowed, it consistently yielded one of the ground states at each minimization trial.

The density of states reconstructed from the hybrid sampling is presented in Fig.~\ref{fig:Wcomparison} alongside the ground truth result, i.e., the statistical distribution of equivalent reconstructions based on the uniform sampling of the exhaustively enumerated states. The comparison shows that the reconstruction based on the hybrid classical-quantum sampling is fully consistent with the ground truth reference. For instance, about half of the data points (4 out of 7 bins) fall within the Q1-Q3 interquartile range.

The result provides a proof-of-concept demonstration of the feasibility of using quantum optimization platforms for intervalled samplings that are sufficiently uniform that accurate reconstructions of the density of states can be achieved. With the constant improvements of quantum platforms, our method could thus enable studying QUBO-mapped physical models with significant speed improvements over classical optimizers.

\section{Conclusions}
We have introduced a general method to compute expectation values of observables by reweighting microstates obtained by minimizing QUBO Hamiltonians of physical models.

QUBO models are natively suited to sample systems with fixed order parameters, e.g.~microstates at a fixed energy of the physical model, since their values can be straightforwardly fixed with quadratic constraints in the QUBO Hamiltonian. Our method enables computing observables in the conjugate ensembles, e.g. canonical averages at fixed temperature. First, we turn the quadratic constraints of the order parameters into quadratic restraints by introducing slack variables. In this way, the QUBO-based sampling can be directed towards finite intervals of the constrained order parameter. Next, we target an overlapping series of intervals that cover the entire range of interest of the order parameter. The gathered states are then processed with a generalized histogram reweighting technique to optimally reconstruct the density of states, which is finally harnessed to compute the sought expectation values in the conjugate ensembles.

The general formulation of our method makes it usable with different conjugate ensembles, as we demonstrated by using the method in two different contexts.

First, we validated the approach for the 2D Ising model, for which the density of states is known exactly. Using a $12\times12$ system, we demonstrated that the method enables a bias-free reconstruction of the entire density of states. By using multiple energy intervals, and sampling negligible fraction of the entire configuration space, we achieved an average relative reconstruction accuracy of order $10^{-2}$ across the 40 orders of magnitudes spanned by the density of states.

Finally, we explored the topological entanglement of a melt of self-assembled semiflexible rings, which is relevant across diverse contexts, from polymer physics to synthetic supramolecular constructs and designed metamaterials. The problem is challenging for conventional sampling methods using real-space representations, and no results have heretofore been established for ring melts about how intra- and inter-ring entanglement vary with the bending rigidity. By leveraging the available efficient mapping to QUBO models and applying our reweighting method to states sampled in multiple intervals of the bending energy, we obtained the knotting and linking probabilities for a broad range of bending rigidities, the conjugate order parameter. We thus established, among other results, that the linking probability is non-monotonic and can be maximized at a suitable value of the bending rigidity. The result establishes, for the first time, that the linking
probability in systems of topologically unrestricted ring polymers has a non-monotonic dependence on bending rigidity. Besides advancing the characterization of dense polymer self-assemblies beyond state-of-the-art real-space sampling methods, the findings suggest new ways to optimize mechanical-bonding in extended supramolecular assemblies, such as Olympic gels.

In both these contexts, the uniform sampling of the ground state manifold of the QUBO Hamiltonians was performed using a parallel tempering scheme on a classical computer. However, quantum optimization platforms are the ideal avenue for our method because they can afford increasing practical speedups as their size and performance continues to improve.
To this end, we used classical-quantum optimizers to sample exhaustively-enumerable ring melts, and demonstrated the feasibility of obtaining accurate reconstructions of the density-of-states for computing canonical averages.

Our scheme can be extended in several directions, both for formulation and applications.
For instance, the method can be generalized to reconstruct the density of states that are functions of several order parameters, $W(\mu_1, \mu_2, ...)$. This would involve equipping the QUBO Hamiltonian with a quadratic restraint for each parameter and generalizing the self-consistent equations for $W$.
In addition, while the powers-of-two linear combination of slack variables provides a natural uniform coverage of the intervals, it may be more efficient to devise other combination schemes designed to counteract the entropic suppression arising from wide dynamic ranges of the density of states. In this case, the self-consistent equations for $W$ would need to be adjusted to take into account the sampling biases introduced {\em ad hoc}. By the same token, data from the intervalled QUBO sampling could be combined with that from unconstrained sampling, performed with QUBO or even conventional methods.

Prospective applications of our method include QUBO-mapped physical systems that call for being treated in the canonical ensemble. For instance, incorporating finite-temperature considerations could enhance the realism of models in soft matter and biological physics, including protein folding, protein design, and RNA secondary structure predictions, which have all been mapped to QUBO models.

\section{Acknowledgments}
We are grateful to Pietro Faccioli, Philipp Hauke, and Guglielmo Mazzola for feedback on the manuscript. We thank CINECA for access to the D-Wave platform under the PRACE programme. This study was funded in part by the European Union - NextGenerationEU, in the framework of the PRIN Project "The Physics of Chromosome Folding" (code: 2022R8YXMR, CUP: G53D23000820006) and by PNRR Mission 4, Component 2, Investment 1.4\_CN\_00000013\_CN-HPC: National Centre for HPC, Big Data and Quantum Computing - spoke 7 (CUP: G93C22000600001). The views and opinions expressed are solely those of the authors and do not necessarily reflect those of the European Union, nor can the European Union be held responsible for them.

\vfill \eject

\section*{Appendix A: Density of states reconstruction for the $L\times L$ Ising system}
\subsection*{QUBO model for the $L\times L$ Ising system}
To perform the QUBO-based reconstruction of $W$ for the 2D Ising system, we introduce the following QUBO-Hamiltonian, which involves site ($\sigma_i$) and edge ($\eta_{ij}$, $\theta_{ij}$) variables on the $L\times L$ lattice, as well as slack variables ($s_k$):
\begin{equation}
    \mathcal{H} = \mathcal{H}_0 + A \mathcal{H}_s = \sum_{\langle ij \rangle}V_{ij} + A \mathcal{H}_s\ ,
    \label{eqn:Isingham}
\end{equation}
where:
\begin{align}
    V_{ij} &= \left( 1 + 2\sigma_i \sigma_j + 2(\sigma_i+\sigma_j) \eta_{ij} \right. \nonumber \\
    &\left.- 4 (\sigma_i+\sigma_j + \eta_{ij}) \theta_{ij} -\sigma_i -\sigma_j - \eta_{ij} + 8 \theta_{ij} \right)  \\
    \mathcal{H}_s &= \left( \sum_{\langle ij \rangle} \eta_{ij} - 2 \bar{n}_{\shortparallel} - 2 \sum_{k=0}^{m-1} 2^k s_k \right)^2 \ .
\end{align}
\noindent In the above expressions, $A$ is a non-negative coefficient, which we set equal to 1, and $\sum_{\langle ij \rangle}$ indicates the summation over distinct pairs of neighboring lattice sites.

The $\sigma_i={0,1}$ binary site variables are in one-to-one correspondence with the up/down spins of the physical Ising model. The $\eta$ binary variables are instead introduced to keep track of parallel and antiparallel neighboring spins. They are tied to the $\sigma$s by an XNOR relation, which is enforced as a quadratic QUBO constraint involving the binary ancilla variables, $\theta$.
This is best seen by setting $A=0$, which reduces the QUBO Hamiltonian to ${\cal H}_0$. In this case, minimizing the Hamiltonian for a given spin configuration $\sigma$ amounts to minimizing the individual quadratic terms $V_{ij}$, yielding the sought XNOR relation:
\begin{equation}
    \eta_{ij}=
\begin{cases}
    1 & \text{if } \sigma_i=\sigma_j,\\
    0              & \text{otherwise.}
\end{cases}
\label{eqn:xnor}
\end{equation}

We note that the expressions for $V_{ij}$ are constructed such that for any set of the $\sigma$ variables -- each mappable to a unique set of physical spins -- the energy-minimizing $\eta$s and $\theta$s yield ${\cal H}_0=0$.
Thus, the degenerate ground states of ${\cal H}_0$ are in one-to-one correspondence with the possible spin configurations and, thanks to the ancilla $\theta$ variables, also with the associated $\eta$s, which represent the parallel or antiparallel alignment of neighboring spins.

\begin{figure}[!ht]
    \centering
\includegraphics[width=0.47\textwidth]{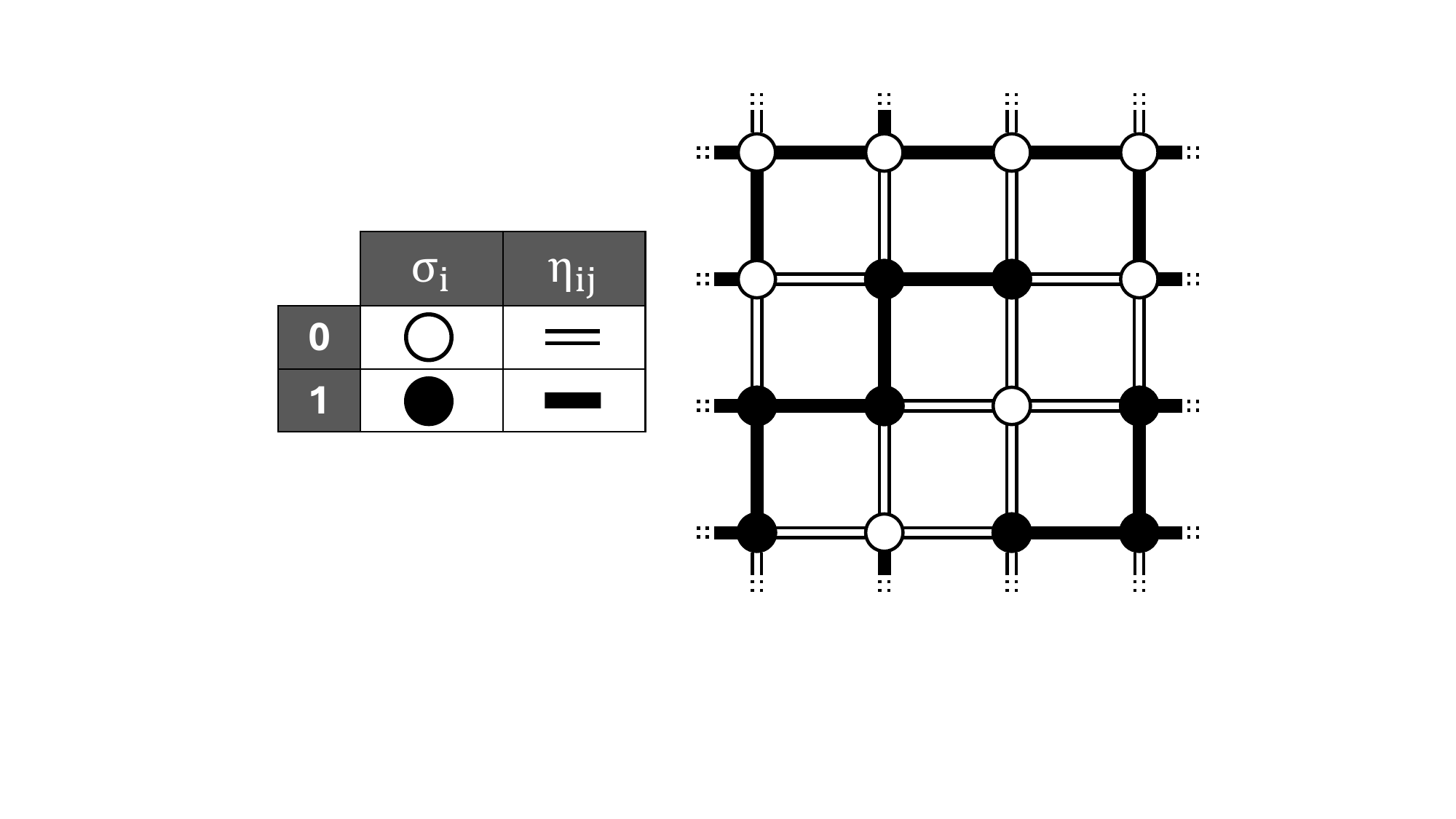}\\
    \caption{\textbf{Example of QUBO encoding for energy-constrained Ising models.}  Left: Symbolic representation of QUBO binary variables associated to up/down physical spins ($\sigma_i$) and parallel/antiparallel spin pairs ($\eta_{ij}$). Right: The represented state is one of the possible ground state solutions of the QUBO Hamiltonian of eq.~\ref{eqn:Isingham} formulated for a $4\times4$ periodic Ising system with ${\cal H}_s$ parameters set to $\bar{n}_{\shortparallel}=8$ and $m=0$. The ground state solutions correspond to Ising spin configurations with $16$ pairs of parallel neighbouring spins, including those wrapping across the periodic boundary conditions (dotted).}
    \label{fig:Ising_notation}
\end{figure}

A schematic representation of the $\sigma$ and $\eta$ variables of a ground state solution of ${\cal H}_0$ in a $4\times 4$ system is shown in Fig.~\ref{fig:Ising_notation}.

The second quadratic term in eq.~\ref{eqn:Isingham}, which involves slack variables, has the same structure as that of eq.~\ref{eqn:energyinterval}. Its minimization ensures that the number of parallel spins, $2n_{\shortparallel} = \sum_{\langle ij \rangle} \eta_{ij}$, falls within the interval $[2\bar{n}_{\shortparallel}, 2\bar{n}_{\shortparallel}+2^{m+1}-2]$.

Thus, by minimizing the total Hamiltonian of Eq.~\ref{eqn:Isingham} the sampling can be targeted at specific intervals of $n_{\shortparallel}$, as needed for reconstructing the density of states $W(n_{\shortparallel})$.

\begin{figure}[!ht]
    \centering
    \includegraphics[width=0.46\textwidth]{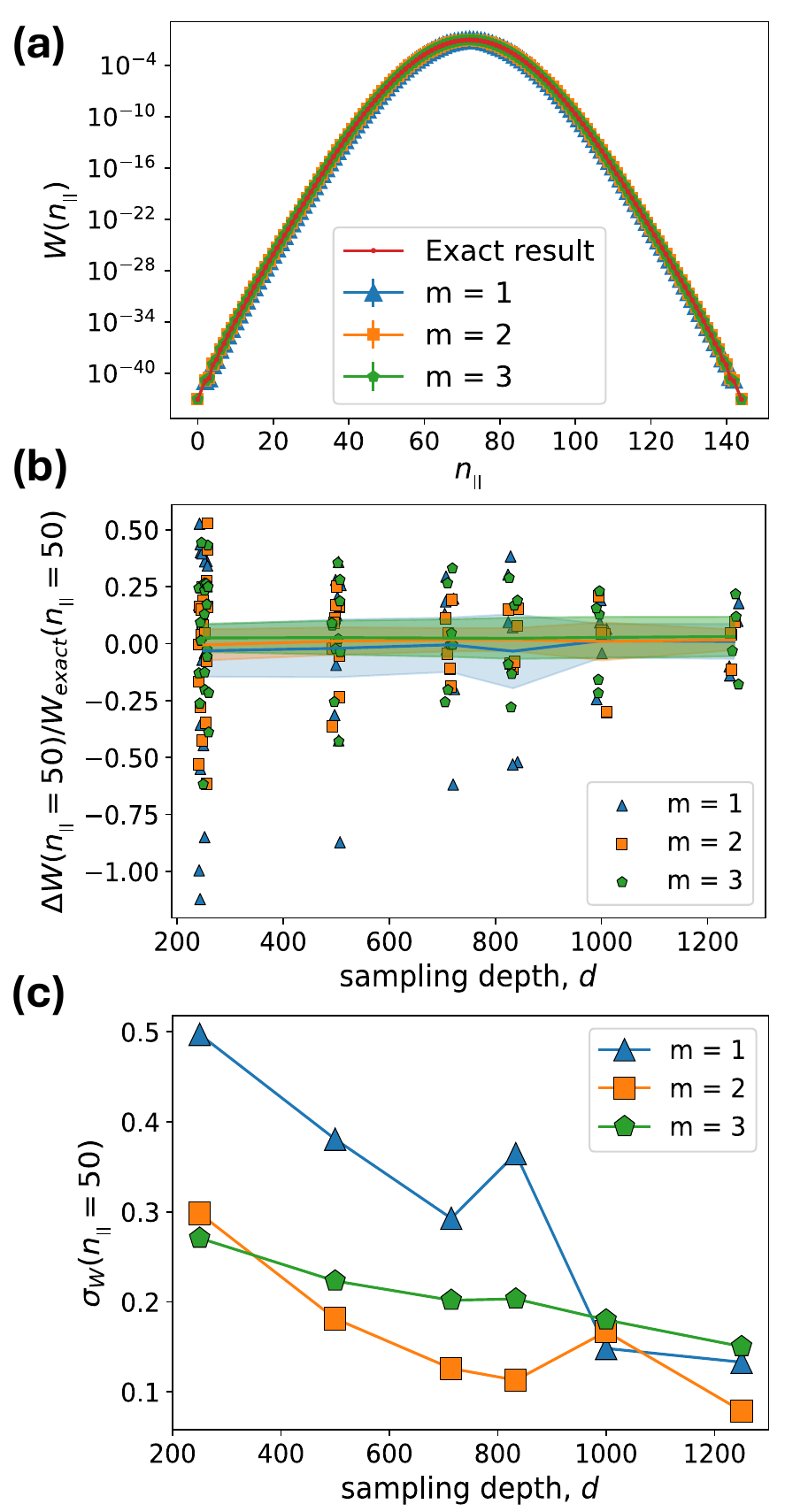}
    \caption{\textbf{Density of states of the $12 \times 12$ Ising model.}
    {\bf (a)} Exact and reconstructed density of states, $W(n_{\shortparallel})$ for a $12 \times 12$ Ising model as a function of the number of parallel neighbouring spins. The data in this and the following panels are color-coded according to the number of slack variables and distinguished by different marker shapes.
    {\bf (b)} Normalized reconstruction error, $\Delta W(n_{\shortparallel}=50)/W_{\rm exact}(n_{\shortparallel}=50)$, for different sampling depths, $d$. The solid lines connect the average values of $\Delta W(n_{\shortparallel}=50)$ computed from independent reconstructions. The data points include jitter along the $x$-axis for visual clarity. Non-overlapping blocks of $d$ samples were used from a population of 5000 independent samples per interval. The shaded bands indicate the standard error of the mean.  {\bf (c)} Statistical error of the reconstructed density of states, $W$, as a function of sampling depth, $d$, for $n_{\shortparallel}=50$.}
    \label{fig:Ising}
\end{figure}

\subsection*{Comparison of exact and reconstructed $W$s}

We considered the $12\times 12$ Ising system ($L=12$) with periodic boundary conditions. The argument $n_{\shortparallel}$ can take on  $L^2-1$ distinct values, corresponding to all integers between $0$ and $144$, except for $1$ and $L^2-1 = 143$. The peak value of $W$ is $\sim 2\times10^{42}$, thus indicating that the "dynamic range" of the density of states spans across more than 40 orders of magnitude.

For the QUBO-based reconstruction, we used multiple intervalled samplings with unit increments in $\bar{n}_{\shortparallel}$.
We considered intervals of width $2$, $4$, and $8$, corresponding to $m=1$, $2$ and $3$ slack variables, respectively.

We covered each of the $\sim 150$ intervals with a sampling depth of $d\sim 5,000$ independent states. The latter were obtained using a classical parallel tempering scheme to minimize the QUBO Hamiltonian (See Section 2 of the Supplementary Material.).

The reconstructed $W$s are compared with the normalized exact one \cite{beale1996exact} in the semi-log plot of Fig.~\ref{fig:Ising}a,
which shows a remarkably consistent agreement throughout the 40 orders of magnitude spanned by $W$.

The reconstructed profiles are visually indistinguishable from the exact one at all $m$s, although there are differences. While $W(n_{\shortparallel})$ can be accurately reconstructed across the entire range of the argument using $m=2$ and $m=3$, this is not the case for $m=1$, where $W$ is not resolved close to boundaries $n_{\shortparallel}=0$ and $n_{\shortparallel}=L^2$. This is because $m=1$ intervals are too short to bridge the gaps at $n_{\shortparallel}=1, L^2-1$.

At the same time, longer and longer intervals are not necessarily beneficial, as they would cover broader dynamic ranges of $W$, thus increasing the sampling depth $d$ required to cover the entire interval. Thus, the optimal choice of $m$ should be made by taking into account the variation of $W$ and the gaps of its argument.

To quantitatively assess the effect of sampling depth $d$ on reconstruction fidelity, we divided the $5,000$ collected states for each interval into $s$ non-overlapping blocks, each comprising $d$ microstates, with $d$ ranging from $200$ to $1250$. Next, we obtained $s$ independent reconstructions, $W_1$, $W_2$, ... , $W_s$, by using respectively only the data from the first block of each interval, then from the second block, etc.

We next considered the pointwise error of the profiles:
\begin{equation}
   \Delta W_i(n_{\shortparallel})= W_i(n_{\shortparallel}) - W_{\rm exact}(n_{\shortparallel})\ ,
\end{equation}
and computed the mean and variance of these errors across the $s$ equivalent and independent reconstructions:
\begin{align}
    \label{eqn:avgDeltaW}
    \langle \Delta W(n_{\shortparallel}) \rangle &=\langle W(n_{\shortparallel}) \rangle - W_{\rm exact}(n)  \nonumber\\
    {\rm Var}\left[ \Delta W (n_{\shortparallel}) \right]&= \langle \Delta W^2(n_{\shortparallel}) \rangle - \langle \Delta W(n_{\shortparallel}) \rangle^2 \\
    &={\rm Var}\left[ W (n_{\shortparallel}) \right] \nonumber
\end{align}
where $\langle . \rangle$ denotes the average over the $s$ block estimates.

Fig.~\ref{fig:Ising}b shows the results of the error analysis for the representative bin $n_{\shortparallel}=50$, corresponding to one of the two midpoints of $\log(W)$. For clarity, the data are normalized to the exact $W$ value, $W_{\rm exact}(n_{\shortparallel}=50)$.
For all considered $m$s, we observe that the $\Delta W_i$ values (data points) are clustered around zero. In fact, their means (solid lines) are compatible with zero within the estimated error on the mean (shaded band), which is equal to:
\begin{equation}
    {\rm Var}\left[ \langle \Delta W (n_{\shortparallel}) \rangle \right] = \frac{{\rm Var}\left[ \Delta W (n_{\shortparallel}) \right]}{\sqrt{s - 1}}
\end{equation}

Note that the results also establish that $\langle W \rangle$ is compatible with $W_{\rm exact}$ within the estimated error. Considering that the latter can be calculated without reference $W_{\rm exact}$ (see eq.~\ref{eqn:avgDeltaW}), we conclude that multiple independent reconstructions of the normalized $W$ allow for computing the associated statistical uncertainty in a reliable and unbiased manner.

In addition, the third panel shows the error on the individual reconstructions,  $\sqrt{{\rm Var}\left[ \Delta W (n_{\shortparallel}) \right]}$, which decreases with the sampling depth for all $m$s.

\section*{Appendix B: melt of self-assembled rings with varying bending rigidity}

\subsection*{QUBO-based sampling of self-assembled ring melts}
We consider ring melts that completely fill cubic lattices of $N$ index sites. A QUBO Hamiltonian for such systems can be formulated in terms of two types of Ising-like binary variables, hereafter indicated as $\Gamma^-$  and $\Gamma^\llcorner$.  A $\Gamma^-_{ij}$ variable is attached to each lattice edge, $ij$, to indicate whether a bond is present (1, active) or absent (0, inactive) between neighboring sites $i$ and $j > i$. Similarly, a $\Gamma^{\llcorner}_{ijk}$ variable is assigned to each corner triplet of sites, $j$ being neighbor to both $i$ and $k > i$. This variable indicates whether the two incident edges $ij$ and $jk$ are both occupied by bonds (1, active) or not (0, inactive).

\begin{figure}[!ht]
    \centering
    \includegraphics[width=0.47\textwidth]{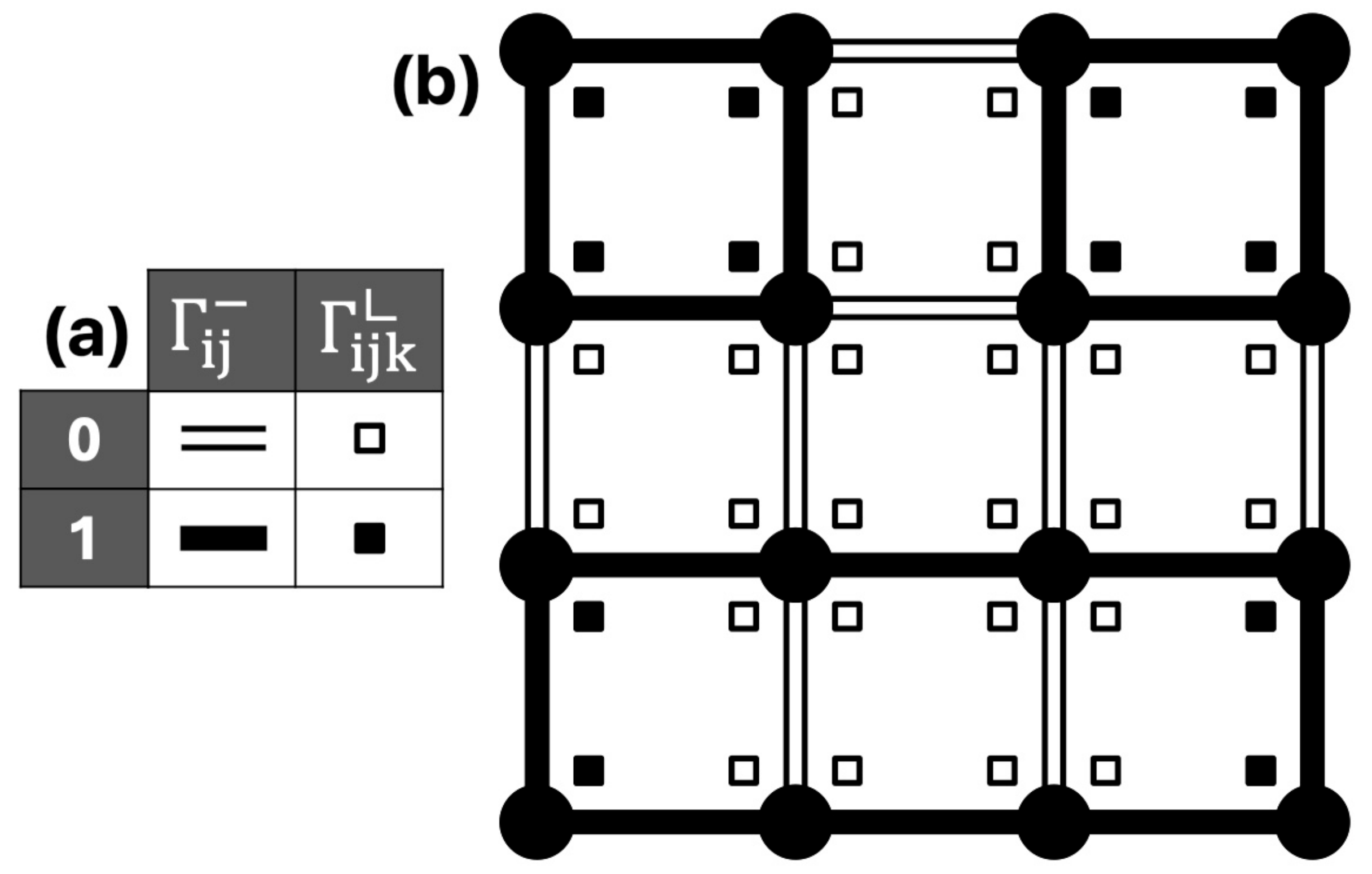}\\
    \caption{\textbf{Example of QUBO encoding for space-filling melts of ring polymers.} {\bf (a)} Symbolic representation of QUBO binary variables corresponding to occupied/empty bonds ($\Gamma_{ij}^{-}$) and corner triplets ($\Gamma_{ijk}^{\llcorner}$). {\bf (b)} The represented configuration is one of the possible ground state solutions of the QUBO Hamiltonian of eq.~\ref{eqn:Hplain} formulated for a $4\times 4$ square lattice and with the setting $N=16$ to impose space-filling. In this example, the total curvature (number of corner turn) is $n_c=12$.}
    \label{fig:Polymer_notation}
\end{figure}

Using the above variables, the QUBO Hamiltonian for the system reads \cite{slongo2023quantum}:
\begin{eqnarray}
&{\cal H}_N= \label{eqn:v1} A_\textrm{b} \left({\sum}_{\langle ij\rangle} \Gamma^-_{ij} -N\right)^2\nonumber + A_{\textrm{c}}{\sum}_{\langle ijk\rangle} {\sum}'_{\langle ljm\rangle} \Gamma^\llcorner_{ijk} \Gamma^\llcorner_{ljm} \\
&+ A_{\textrm{bc}} {\sum}_{\langle ijk\rangle} \left[3 \Gamma^\llcorner_{ijk} + \Gamma^-_{ij}\Gamma^-_{jk}  -2 \Gamma^\llcorner_{ijk} (\Gamma^-_{ij}+ \Gamma^-_{jk})\right]
\label{eqn:Hplain}
\end{eqnarray}
\noindent where the $A$ coefficients are non-negative, $\sum_{\langle i j \rangle}$ and $\sum_{\langle i j k\rangle}$ indicate summations over distinct neighboring pairs and triplets of lattice sites, respectively. The prime indicates the restriction over inequivalent triplets, $\langle ijk \rangle \neq \langle ljm\rangle$.

The first quadratic term is minimized when the total number of bonds (active edges) is equal to the number of lattice sites $N$, as required by the space-filling condition of the ring melt. 
The second quadratic term is minimized when no branching is present because it penalizes cases where three or more bonds meet at the same lattice site. Combining this constraint with the first one, i.e., that the number of active bonds is equal to the number of sites, implies that each site has exactly two incident bonds and is, therefore, part of a closed chain. The third term is a quadratic constraint that enforces the consistency of the $\Gamma^{-}$ and $\Gamma^{\llcorner}$ variables. In fact, this term is minimized if and only if the active corners of $\Gamma^{\llcorner}$ are compatible with the active bonds of $\Gamma^{-}$, i.e.,
\begin{equation}
    \Gamma^{\llcorner}_{ijk}=
\begin{cases}
    1, & \text{if } \Gamma^{-}_{ij}=\Gamma^{-}_{jk}=1\\
    0,              & \text{otherwise.}
\end{cases}
\end{equation}

Thus, minimizing all three terms simultaneously yields a binary encoding of self-assembled polymers that satisfy the physical requirements of being space-filling, self-avoiding, and exclusively consisting of closed chains. Notice that the number of closed chains is not fixed and is determined by the self-assembly combinatorics.

The ground states of the QUBO Hamiltonian of eq.~\ref{eqn:Hplain}, which by construction correspond to ${\cal H}_N=0$, are thus in one-to-one correspondence with the configurations of maximally-dense melts of rings.

A schematic representation of the $\Gamma^-$  and $\Gamma^\llcorner$ of a ground state solution of ${\cal H}_N$ is shown in Fig.~\ref{fig:Polymer_notation}. For clarity, the illustrated case is for a two-dimensional $4 \times 4$ lattice.

\subsection*{QUBO-based sampling for a bending energy interval}

As noted in connection with eq.~\ref{eqn:Observable_final}, a natural parameter for profiling the density of states $W$ is the total number of corner turns, $n_c$. This quantity is proportional to the total curvature and, hence, is the conjugate variable of the bending rigidity. In the ground state manifold ${\cal H}_N$, $n_c$ can be directly computed from the number of active $\Gamma^{\llcorner}$ variables in the ground states of ${\cal H}_N$:

\begin{equation}
    n_c = \sum_{\langle ijk \rangle} \Gamma^\llcorner_{ijk} \ .
\end{equation}

The intervalled sampling required to reconstruct $W(n_c)$ can thus be achieved by adding to ${\cal H}_N$ a quadratic term involving a set of slack variables $s_0$, $s_1$, $\dots$ , $s_{m-1}$ and proportional to:
\begin{equation}
    {\cal H}_s=\left(\sum_{\langle ijk \rangle} \Gamma^\llcorner_{ijk} - \bar{n}_c - \sum_{k=0}^{m-1} 2^k s_k \right)^2 \ .
\end{equation}
In fact, minimizing ${\cal H}_N + A {\cal H}_s$ with $A>0$ allows for sampling states in the interval $\bar{n}_c \leq n_c \leq \bar{n}_c + 2^m -1$.

Repeating the sampling procedure for overlapping intervals covering the entire range of $n_c$, and applying the reconstruction method of Sec.\ref{sec:2B}, one can obtain the full profile of the density of states $W$.

\subsection*{Ring melt composition versus bending rigidity}
The composition of the self-assembled rings melts is conveniently characterized in terms of the average number of self-assembled rings at fixed temperature and bending rigidity,  $\langle N_{\rm rings} \rangle_{\beta \kappa_b}$. We recall that the number of chains in self-assembling polymer systems can be controlled with extrinsic design parameters, such as monomer density, and intrinsic ones, such as the monomers' bonding volume \cite{sciortino2016basic}. In our case, the former is fixed by the space-filling conditions, while the latter is varied via the bending rigidity.

We computed $\langle N_{\rm rings} \rangle_{\beta \kappa_b}$ using eq.~\ref{eqn:Observable_final} after identifying $\langle \mathcal{O}  \rangle_{n_c}$ with the average number of rings per each admissible value of $n_c$, $\langle N_{\rm rings} \rangle_{n_c}$.

\begin{figure}[!ht]
    \centering
   \includegraphics[width=0.4\textwidth]{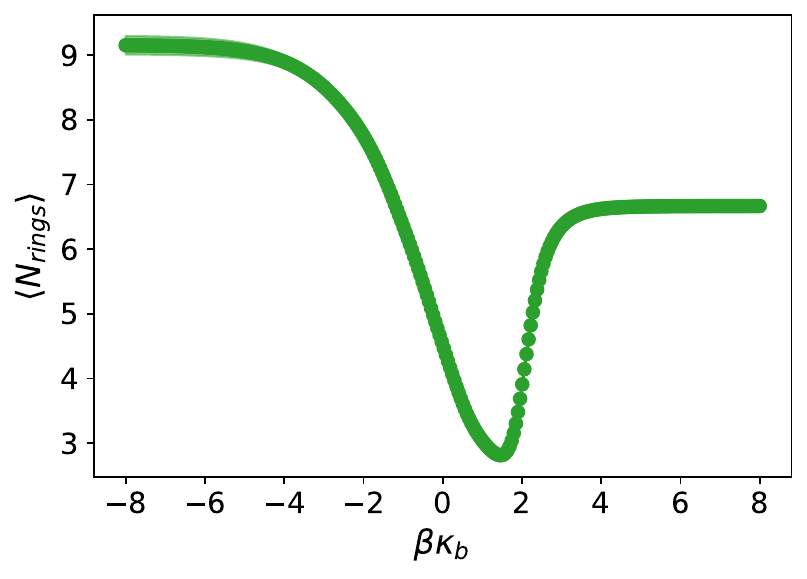}
    \caption{{\bf Average number of self-assembled rings as a function of the reduced bending rigidity $\beta \kappa_b$.} The results are for a canonically equilibrated ring melt filling a $5 \times 5 \times 4$ cubic lattice. The data points represent the average $P_{\rm link}$ computed from $4$ reconstructions and the shaded bands indicate the standard error of the mean. The considered range of $\beta \kappa_b$ spans from $-8$ to $+8$, i.e. from strongly favoured to strongly suppressed corner turns. Across this range, the average number of rings is non-monotonic, having a minimum at $\beta \kappa_b \sim 1.5$.}
    \label{fig:Nring}
\end{figure}

Fig.~\ref{fig:Nring} illustrates the resulting profile of $\langle N_{\rm rings} \rangle_{\beta \kappa_b}$, obtained by averaging five independent reconstructions. The shaded band denotes the estimated statistical error on the average profile, which is typically smaller than $0.6$\% except for $\beta \kappa_b < 4$.

For $\beta \kappa_b > 4$, $\langle N_{\rm rings} \rangle_{\beta \kappa_b \gg 0}$ plateaus at about $6.5$. In this regime of large bending rigidity, typical configurations of the ring melt correspond to nested stacked rings, which typically involves approximately $7$ rings on $5\times5\times4$ space-filled lattice, as illustrated in Fig.~\ref{fig:linkingprobability}e.

For $\beta \kappa_b < -4$, instead, $\langle N_{\rm rings} \rangle$ plateaus to about 9. This larger asymptotic value reflects the fact that, when chain turns are favored, the rings are smaller and hence more numerous, as shown in Fig.~\ref{fig:linkingprobability}b.

Fig.~\ref{fig:Nring} shows that in between two asymptotic values $\langle N_{\rm rings} \rangle_{\beta \kappa_b}$ is minimum for $\beta\kappa_b \sim 1.7$, where it is slightly smaller than 3. A typical configuration for such extremal case is shown in Fig.~\ref{fig:linkingprobability}g-h.

\subsection*{Knotting probability versus bending rigidity}
To characterize the intra-ring entanglement, we considered the knotting probability, defined as the probability that individual rings in the melt are knotted, $P_{\rm knot}$. To compute the knotting dependence on $\beta \kappa_b$, we used eq.~\ref{eqn:Observable_final} after identifying  $\langle O \rangle_{n_c}$ in  with  $P_{\rm knot} \left( n_c \right)$.

\begin{figure}[!h]
    \centering
   \includegraphics[width=0.4\textwidth]{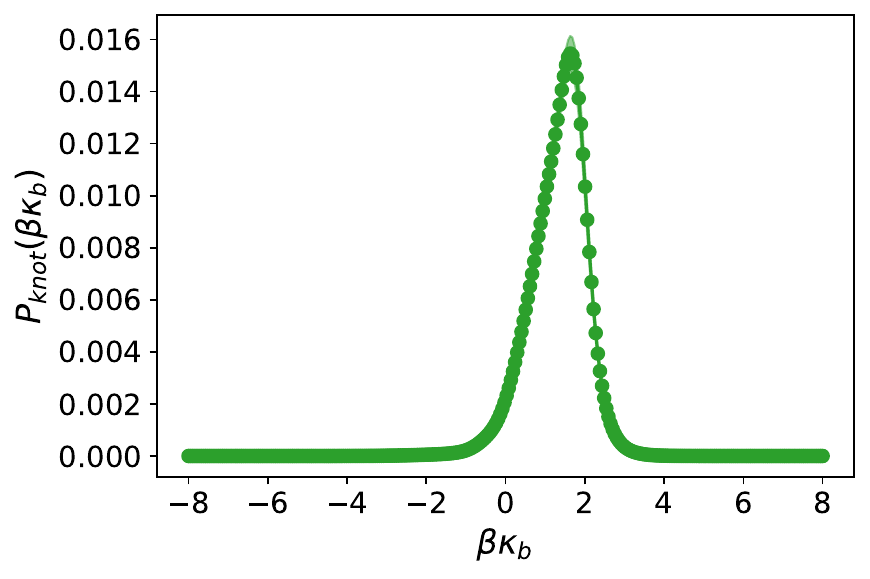}
    \caption{{\bf Equilibrium knotting probability of melts of self-assembled rings as a function of the reduced bending rigidity $\beta \kappa_b$.} Knotting probability $P_{knot}$ for a canonically-equilibrated ring melt filling a $5 \times 5 \times 4$ cubic lattice. The data points represent the average $P_{knot}$ computed from $4$ reconstructions. The associated errors, indicated by the shaded band, are typically smaller than the symbol size, except near the peak. The considered range of $\beta \kappa_b$ spans from $-8$ to $+8$, i.e. from strongly favoured to strongly suppressed corner turns. Across this range, the knotting probability is non-monotonic, peaking at $\beta \kappa_b \sim 1.6$.}
    \label{fig:knotting}
\end{figure}

The resulting $P_{\rm knot}$ vs $\beta \kappa_b$ curve is shown in Fig.~\ref{fig:knotting}. The curve approaches zero for large $\beta |\kappa_b|$, irrespective of the sign of the bending rigidity.

In the limit $\beta \kappa_b > 4$, the vanishing knotting probability is due to the mentioned nested columnar structures. Since the stacked rings are planar, they are necessarily unknotted, too.
Instead, in the opposite situation ($\beta \kappa_b < -4$), the microscopic basis for the vanishing $P_{\rm knot}$ is fundamentally different. In this regime, $n_c$ is largest, corresponding to a $\pi/2$ turn at each lattice site. These tightly wound rings do not leave openings that can be threaded by the other parts of the chain, thereby preventing knot formation, see Fig.~\ref{fig:linkingprobability}b.

Between these two limits, $P_{\rm knot}$ is maximum for $\beta \kappa_b \sim 2$. In this regard, we recall that various systems of isolated rings have been shown to have a non-monotonic knotting probability as a function of bending rigidity, from lattice polymer models \cite{orlandini2005entanglement}, to off-lattice open \cite{virnau2013influence} and closed chains \cite{coronel2017non}. Our results demonstrate that this result, previously established only for isolated chains, equally applies to polymer melts, too.

\subsection*{Supplementary Material available}
The provided supplementary material file contains: further methodological and algorithmic details about solving the self-consistent equations for $W$, the use of parallel tempering schemes for minimizing QUBO Hamiltonians and additional results regarding ring melts.
\bibliography{bibliography}

\end{document}


\title{
{\LARGE Supplementary Material for}
\\[0.5cm]
{\Large Computing canonical averages with quantum and classical optimizers: Thermodynamic reweighting for QUBO models of physical systems}}

\author{Francesco Slongo}
\affiliation{Scuola Internazionale Superiore di Studi Avanzati (SISSA), Via Bonomea 265, I-34136 Trieste, Italy.}
\author{Cristian Micheletti}
\affiliation{Scuola Internazionale Superiore di Studi Avanzati (SISSA), Via Bonomea 265, I-34136 Trieste, Italy.}

\date{\today}

\maketitle

\renewcommand{\theequation}{S\arabic{equation}}
\setcounter{equation}{0}
\renewcommand{\thefigure}{S\arabic{figure}}
\setcounter{figure}{0}
\renewcommand{\thetable}{S\arabic{table}}
\setcounter{table}{0}

\section{Solution of the self-consistent equations for $W$}

We adopted an iterative procedure to solve the self-consistent relation for $W$ of Eq.~(10) of the main text, which we rewrite here more conveniently as:
\begin{equation}
   W(E) =\frac{\sum_j^\prime \frac{n_j(E)}{g_j \left(1 - \frac{W(E)}{Z_j} \right)}}{\sum_k^\prime \frac{N_k}{Z_k g_k \left(1 - \frac{W(E)}{Z_k} \right)}} = \frac{\sum_j^\prime \frac{n_j(E)\, Z_j}{g_j \left(Z_j - W(E)\right)}}{\sum_k^\prime \frac{N_k}{g_k \left(Z_k - W(E)\right)}} \ ,
\label{eqn:final}
\end{equation}
\noindent where:
\begin{equation}
    Z_j=\sum_{E_{\rm min}(j) \le E \le E_{\rm max}(j)} W(E)\ ,
    \label{eqn:Z_definition}
\end{equation}
\noindent and the $W$ array is subject to the normalization condition:
\begin{equation}
    \sum_E W(E) = 1 \ .
    \label{eqn:normalization_condition}
\end{equation}
\noindent The prime in Eq.~\ref{eqn:final} indicates that the sum is restricted to intervals that include $E$.

In the following we shall use a superscript $(i)$ to indicate the $W$ and $Z$ arrays at the $i$th iteration step, i.e., $W^{(i)}(E)$ and $Z^{(i)}_j$. Additionally, we shall assume that $W^{(i)}(E)=0$ in correspondence of energy bins where no states have been observed.

\subsection{Iterative procedure}

With the above proviso, the density of states is initialized to a trial profile $W^{(0)}(E)$, such as the one obtained with the procedures discussed in subsections C and D. The $j$th element of the corresponding $Z$ array,  $Z^{(0)}_j$, is then calculated from Eq.~\ref{eqn:Z_definition} using $W^{(0)}(E)$ in place of $W(E)$.

At the $i$th iteration step the $W$ array elements are updated with a relaxation procedure based on Eq.~\ref{eqn:final}:

\begin{equation}
    W(E)^{(i+1)} = (1-\alpha)\, W^{(i)}(E) + \alpha \, \frac{\sum_j^\prime \frac{n_j(E)\, Z_j^{(i)}}{g_j \left(Z_j^{(i)} - W^{(i)}(E) \right)}}{\sum_k^\prime \frac{N_k}{g_k \left(Z_k^{(i)}  - W^{(i)}(E) \right)}} \ .
  \label{eqn:iterW}
\end{equation}
\noindent where the mixing coefficient $\alpha \in (0:1)$ should be small enough to ensure convergence.

Next, the newly computed $W(E)^{(i+1)}$ are rescaled to satisfy the normalization condition of Eq.~\ref{eqn:normalization_condition}.

Finally, the updated $Z$ array is computed:
\begin{equation}
    Z_j^{(i+1)}=\sum_{E_{\rm min}(j) \le E \le E_{\rm max}(j)} W^{(i+1)}(E)\ ,
    \label{eqn:iterZ}
\end{equation}
\noindent The maximum usable value of the mixing parameter $\alpha$ depends on the number of energy bins in the intervals and the dynamic range of the $W$ elements in the intervals, as discussed below. In practical implementations, we found that the convergence rate of the procedure is improved by nesting the updates of Eqs.~\ref{eqn:iterW} and \ref{eqn:iterZ} within a damped update of an auxiliary $Z$ array, see subsection D.

\subsection{Stability of the fixed point solution}

We now discussed the stability of the iterative procedure and show that it has a fixed point attractor. We shall indicate the fixed point solutions of Eqs.~\ref{eqn:final}--\ref{eqn:normalization_condition} with a $*$ superscript, i.e.~$W^*$ and $Z^*$. In addition, we shall set all $g$s equal to 1 for clarity, equivalent to assuming that histogrammed data have been resampled with a stride sufficiently large to remove correlations.

Consider a perturbed density of states, $\widetilde{W}$, obtained by modifying the fixed point solution in correspondence of a single energy value, $\bar{E}$,
\begin{equation}
\widetilde{W}(E)=
\begin{cases}
W^*(E) + \Delta W & \text{for } E = \bar{E}, \\
W^*(E)  & \text{otherwise},
\end{cases}
\end{equation}
so that
\begin{equation}
\widetilde{Z}_j=
\begin{cases}
Z_j^* + \Delta W & \text{if $\bar{E}$ is in the $j$th energy interval,} \\
Z_j^* & \text{otherwise.}
\end{cases}
\end{equation}

Substitution into the right-hand side of Eq.~(\ref{eqn:iterW}) yields the updated array $\widetilde{W}^\prime$,
\begin{equation}
\widetilde{W}^\prime(E) = (1-\alpha)\, W^{*}(E) + \alpha \, \frac{\sum_j^\prime \frac{n_j(E) \widetilde{Z}_j}{\widetilde{Z}_j-\widetilde{W}(E)}}{\sum_k^\prime \frac{N_k}{\widetilde{Z}_k -\widetilde{W}(\bar{E})}}\ .
\label{eqn:perturb1}
\end{equation}

{\bf Case $E=\bar{E}$.}
We first specialize Eq.~\ref{eqn:perturb1} to the case $E=\bar{E}$. We note that $\widetilde{Z}_k -\widetilde{W}(\bar{E}) = {Z}_k^* -W^*(\bar{E})$ for all intervals that include $\bar{E}$. Additionally, the term $n_j(\bar{E})\widetilde{Z}_j$ in the numerator can be written as $n_j(\bar{E}){Z}_j^* + n_j(\bar{E}) \Delta W$. Therefore, Eq.~\ref{eqn:perturb1} can be rewritten as:
\begin{equation}
\begin{array}{ll}
\widetilde{W}^\prime(\bar{E}) &= W^{*}(\bar{E}) + \alpha \frac{\sum_j^\prime \frac{n_j(\bar{E})}{Z_j^*-W^*(\bar{E})}}{\sum_k^\prime \frac{N_k}{Z_k^* -W^*(\bar{E})}}\,  \Delta W \\
&\\
&= W^{*}(\bar{E}) + \alpha \, A \, \Delta W \ ,
\end{array}
\label{eqn:perturb2}
\end{equation}
\noindent where the non-negative coefficient $A$ is $\leq 1$ because $n_j \leq N_j$.

{\bf Case $E \not=\bar{E}$.}
The update via Eq.~\ref{eqn:perturb1} will generally propagate the initial pointwise perturbation to the neighborhood of $\bar{E}$, and specifically to the $W$s of energy levels spanned by intervals that include $\bar{E}$.
Thus, we shall now consider $E \not=\bar{E}$.
In this case, for intervals that include $\bar{E}$, we have $\widetilde{Z}_j -\widetilde{W}(\bar{E}) = {Z}_j^* -W^*(\bar{E})-\Delta W$, otherwise $\widetilde{Z}_j -\widetilde{W}(\bar{E}) = {Z}_j^* -W^*(\bar{E})$.

Taking this into account and expanding the right-hand side of Eq.~\ref{eqn:perturb1} to linear order in $\Delta W$, one obtains:
\begin{equation}
\widetilde{W}^\prime(E)= W^*(E) + \alpha (B_1 - B_2) \Delta W
\label{eqn:b1b2}
\end{equation}
where the non-negative coefficients $B_1$ and $B_2$ are:
\begin{align}
B_1=&\frac{\sum_{j}^{\prime\prime}  \frac{n_j W^*(E)}{\left(Z^*_j - W^*(E)\right)^2}}{\sum_{k}^\prime \frac{N_k}{Z^*_k - W^*(E)}}\nonumber\\
\null\\
B_2=& W^*(E) \frac{\sum_{l}^{\prime \prime} \frac{N_l}{\left(Z^*_l - W^*(E)\right)^2}}{\sum_{k}^\prime \frac{N_k}{Z^*_k - W^*(E)}}\ ,\nonumber
\end{align}
\noindent with the double prime indicating that the sums are restricted to intervals that include $\bar{E}$.

Considering that: (i) $0 \le W(\bar{E}) \le Z_j - W^*(E)$, (ii) $W^*(E) \le Z_j$, and using the identity of Eq.~\ref{eqn:final} for the fixed point solution, one has the following upper bounds:
\begin{align}
B_1 &\le \frac{1}{W^*(\bar{E})} \frac{\sum_{j}^{\prime}  \frac{n_j Z_j^*}{Z^*_j - W^*(E)}}{\sum_{k}^\prime \frac{N_k}{Z^*_k - W^*(E)}} = \frac{W^*(E)}{W^*(\bar{E})} \nonumber\\
\null\\
B_2  &\le \frac{W^*(E)}{W^*(\bar{E})} \frac{\sum_{l}^{\prime \prime} \frac{N_l}{Z^*_l - W^*(E)}}{\sum_{k}^\prime \frac{N_k}{Z^*_k - W^*(E)}} \le \frac{W^*(E)}{W^*(\bar{E})}\ . \nonumber
\end{align}

We thus obtain the following bound for the perturbation propagated by the update:
\begin{equation}
\| \widetilde{W}^\prime(E) - \widetilde{W}(E)\| \le 2 \alpha \frac{W^*(E)}{W^*(\bar{E})} \Delta W\ .
\label{eqn:perturb3}
\end{equation}

Combining expressions~\ref{eqn:perturb2} and \ref{eqn:perturb3}, we have that the overall perturbation of the updated array has a norm smaller than the initial one as long as the relaxation parameter $\alpha$ is sufficiently small compared to the largest ratio of pairs of $W^*$s in the same intervals. To leading order, for non-pointwise perturbations, the updated entries of $\widetilde{W}^\prime$ will differ from those of $W^*$ by the linear combination of terms analogous to those of Eqs.~\ref{eqn:perturb2} and ~\ref{eqn:b1b2}. Because such terms are bound, and their number is no larger than the preassigned number of bins per interval, a sufficiently small - though finite - value of $\alpha$ will ensure that the norm of the perturbation decays at each iteration step. The fixed point solution is thus an attractor for the iterative procedure.

\begin{figure*}[!t]
  \centering
\begin{tikzpicture}
  \tikzset{
    squarebox/.style={
      draw, rectangle,
      align=center,
      inner sep=1pt,
      outer sep=0pt,
      minimum width=1.3cm,
      minimum height=8mm
    },
    skipsquare/.style={
      rectangle, draw=none,
      align=center,
      inner sep=1pt,
      outer sep=0pt,
       minimum width=1.2cm,
      minimum height=8mm
    }
  }
  \node[squarebox]                         (mid1)    {$n_1(1)$};
  \node[squarebox,  anchor=west]           (mid2)    at (mid1.east)   {$n_1(2)$};
  \node[skipsquare, anchor=west]           (mid3)    at (mid2.east)   {...};
  \node[squarebox,  anchor=west]           (mid4)    at (mid3.east)   {$n_1(l-1)$};
  \node[squarebox,  anchor=west]           (mid5top) at (mid4.east)   {$n_1(l)$};
  \node[squarebox,  anchor=north]          (bot5)    at (mid5top.south) {$n_2(l)$};
  \node[squarebox,  anchor=west]           (bot6)    at (bot5.east)     {$n_2(l+1)$};
  \node[skipsquare,  anchor=west]           (bot7)    at (bot6.east)     {...};
  \node[squarebox,  anchor=west]           (bot8)    at (bot7.east)     {$n_2(2l-2)$};
  \node[squarebox,  anchor=west]           (bot9)    at (bot8.east)     {$n_2(2l-1)$};
  \node[squarebox, anchor=south, yshift=5mm] (top1) at (mid1.north) {$E_1$};
  \node[squarebox, anchor=west]             (top2)  at (top1.east)  {$E_2$};
  \node[skipsquare, anchor=west]            (top3)  at (top2.east)  {...};
  \node[skipsquare, anchor=west]            (top4)  at (top3.east)  {};
  \node[squarebox, anchor=west, yshift=13mm]            (top5)  at (mid4.east)  {$E_l$};
  \node[skipsquare, anchor=west]            (top6)  at (top5.east)  {};
  \node[skipsquare, anchor=west]            (top7)  at (top6.east)  {...};
  \node[squarebox,  anchor=west, yshift=21mm]            (top8)  at (bot7.east)  {\phantom{xx}$E_{2l-2}$\phantom{cc}};
  \node[squarebox,  anchor=west]            (top9)  at (top8.east)  {\phantom{xx}$E_{2l-1}$\phantom{ci}};
  \coordinate (labelX) at ($(mid1.west) + (-35mm,0)$);
  \node[anchor=base west, align=left,yshift=-3mm] at (labelX |- top1.base)
    {energy levels (bins)\\$\null$};

  \node[anchor=base west, align=left,yshift=0mm] at (labelX |- mid1.base)
    {interval 1 histogram};

  \node[anchor=base west, align=left,yshift=0mm] at (labelX |- bot5.base)
    {interval 2 histogram};

\end{tikzpicture}
  \caption{Schematic representation of a system with $2l-1$ energy levels (histogram bins). The energy range (top row) is covered by two intervals of $l$ bins each, overlapping at their edges (middle and bottom row). The notation $n_j(i)$ indicates the population of $i$th energy bin sampled in the $j$th histogram.}
  \label{fig:mySchema}
\end{figure*}

\subsection{Approximate solution for initialization}
\label{sec:approx}
We now derive an approximate solution to the self-consistent equations \ref{eqn:final} - \ref{eqn:normalization_condition}. Given its simplicity, the approximate solution can be straightforwardly used as the starting point of the iterative scheme, making convergence significantly faster than, e.g., starting from an initially flat $W$ profile.

The approximation hinges on the assumption that the individual $W$ elements are much smaller than the $Z$s of the intervals containing them. In this limit, the self-consistent equation \ref{eqn:final} reduces to:
\begin{equation}
\label{eq:poisson}
W(E) =\frac{\sum_j^\prime \frac{n_j(E)}{g_j}}{\sum_k^\prime \frac{N_k}{Z_k\, g_k}}
\end{equation}
\noindent Setting all $g$'s equal to 1 for simplicity, and rearranging the terms, the equation can be rewritten more conveniently as:
\begin{equation}
\frac{\sum_j^\prime n_j(E)}{W(E)} ={\sum_k^\prime \frac{N_k}{Z_k}}.
\end{equation}

\subsubsection{Analytical solution}
\label{sec:solvedpoisson}
We further assume that the energy range of interest is spanned by intervals (hence histograms) that overlap only at the edge bins. The requirement of histograms overlapping at the edges is not restrictive in applicative contexts, in that multiple histogram coverages can always be selectively resampled {\em a posteriori} to meet this condition.

We focus on the case of two intervals of equal length $l$, covering $2l-1$ energy bins in total, see schematic in Fig.~\ref{fig:mySchema}. The same derivation scheme can be used for an arbitrary number of intervals of possibly different lengths as long as they overlap only at the edges, yielding the same final expressions obtained here.

The self-consistent equations for the first $l-1$ energy levels read:
\begin{eqnarray}
\left\{
\begin{aligned}
&\frac{n_1(1)}{W(1)}=\frac{N_1}{Z_1}\\
&\frac{n_1(2)}{W(2)}=\frac{N_1}{Z_1}\\
& \ldots\\
&\frac{n_1(l-1)}{W(l-1)}=\frac{N_1}{Z_1}
\end{aligned}
\right.
\end{eqnarray}
\noindent where $N_1=n_1(1)+n_1(2)+ \cdots +n_1(l)$, and $Z_1=W(1)+W(2)+ \cdots +W(l)$.

Because the right-hand sides of the equations are identical, so are their left-hand sides, and thus the set can be rewritten as:
\begin{eqnarray}
\left\{
\begin{aligned}
&\frac{n_1(1)}{W(1)}=\frac{N_1}{Z1}\\
&\frac{W(1)}{W(2)}=\frac{n_1(1)}{n_1(2)}\\
& \ldots\\
&\frac{W(l-2)}{W(l-1)}=\frac{n_1(l-2)}{n_1(l-1)}
\end{aligned}
\right.
\end{eqnarray}
In addition, because all ratios $n_1(i)/W(i)$ are identical and equal to $N_1/Z_1$ for $i=1, \ldots, l-1$, the first equation , $n_1(1)/W(1)=(n_1(1)+n_1(2)+ \cdots +n_1(l))/(W(1)+W(2)+ \cdots +W(l))$ implies that $n_1(l)/W(l)$ is equal to the other ratio, too.
Thus, the first $l-1$ equations can be recasted to fix the ratios of all pairs of consecutive $W$s in the first interval:
\begin{equation}
\frac{W(i)}{W(i+1)}=\frac{n_1(i)}{n_1(i+1)} \quad \quad \mbox{for } i=1, 2, \ldots l-1.
\label{eqn:ratios}
\end{equation}

\noindent We now move to the remaining set of equations:
\begin{eqnarray}
\left\{
\begin{aligned}
&\frac{n_1(l) + n_2(l)}{W(l)}=\frac{N_1}{Z_1} + \frac{N_2}{Z_2}\\
&\frac{n_2(l+1)}{W(l+1)}=\frac{N_2}{Z_2}\\
& \ldots\\
&\frac{n_2(2l-1)}{W(2l-1)}=\frac{N_2}{Z_2}
\end{aligned}
\right.
\end{eqnarray}
\noindent where $N_2=n_2(l)+n_2(l+1)+ \cdots +n_2(2l-1)$, $Z_1=W(1)+W(2)+ \cdots +W(l)$, and $Z_2=W(l)+W(l+1)+ \cdots +W(2l-1)$.
Substituting $n_1(l)/W(l)=N_1/Z_1$ in the first equation, yields
\begin{eqnarray}
\left\{
\begin{aligned}
&\frac{n_2(l)}{W(l)}=\frac{N_2}{Z_2}\\
&\frac{n_2(l+1)}{W(l+1)}=\frac{N_2}{Z_2}\\
& \ldots\\
&\frac{n_2(2l-1)}{W(2l-1)}=\frac{N_2}{Z_2}
\end{aligned}
\right.
\end{eqnarray}
\noindent Proceeding as before on the first $l-1$ equations, fixes the ratios of the remainder pairs of consecutive $W$s in exactly the same form of Eq.~\ref{eqn:ratios} also for $i=l, l+1, \ldots l-2$.

\noindent Such relationship automatically satisfies the remaining equation $n_2(2l-1)/W(2l-1)=N_2/Z_2$, which is thus redundant. This is consistent with the fact that, based on the sampled histograms, the $W$ profile can be reconstructed only up to a multiplicative constant, which is fixed by the normalization condition of Eq.~\ref{eqn:normalization_condition}.

To conclude, a deterministic approximate solution of the self-consistent equation can be obtained by considering
intervals overlapping only at the edges, in which case the entire non-normalized profile of $W$ can be constructed from the ratio of consecutive entries, which is equal to the ratio of corresponding populations in the histogram that includes both energy bins.

\subsection{Algorithmic implementation}
\label{sec:algo}
Below is the pseudocode of the algorithm used for the iterative solution of the self-consistent equations, employing nested relaxations. We typically set $N_{\text{cycles}}=100$, $\epsilon = 10^{-15}$,  $N_\text{iter}=50,000$.

\begin{algorithm}
\caption{Pseudocode of the iterative method.}
\begin{algorithmic}[1]
\State Initialize $\vec{W}^{(0)}$
\State Compute $\vec{Z}^{(0)}$ from $\vec{W}^{(0)}$ \Comment{Eq.~\ref{eqn:iterZ}}
\State $\vec{Z}^{\text{damp}} \gets \vec{Z}^{(0)}$
    \State $i \gets 1$
\For{$\text{cycle} \gets 1$ \textbf{to} $N_{\text{cycles}}$}
    \Repeat
        \State $c \gets 1$
        \ForAll{$E$ in energy range}
            \State $num \gets 0$
            \State $denom \gets 0$
            \ForAll{$j$ in histograms that include $E$}
                \State $r \gets 1 - W^{(i-1)}(E)/Z_j^{(i-1)}$
                \State $num \gets num + {n_j(E)}/({g_j\cdot r })$
                \State $den \gets den + {N_j}/({Z_j^{\text{damp}}\cdot g_j \cdot r})$
                \EndFor
            \State $W^{(i)}(E) \gets \alpha \cdot \dfrac{num}{den} + (1 - \alpha)\cdot W^{(i-1)}(E)$
        \EndFor
        \State Normalize $\vec{W}^{(i)}$ \Comment{Eq.~\ref{eqn:normalization_condition}}
        \State Compute $\vec{Z}^{(i)}$ from $\vec{W}^{(i)}$ \Comment{Eq.~\ref{eqn:iterZ}}
        \State $i \gets i + 1$
        \State $c \gets c + 1$
        \State $\delta \gets 0$
        \ForAll{$E$ in energy range}
            \State $\delta \gets \delta + \frac{\left|W^{(i)}(E) - W^{(i-1)}(E)\right|}{W^{(i-1)}(E)}$
        \EndFor
    \Until{$\delta < \epsilon$ \textbf{or} $c \geq N_{\text{iter}}$}
    \State $\vec{Z}^{\text{damp}} \gets \alpha \vec{Z}^{(i)} + (1 - \alpha) \vec{Z}^{\text{damp}}$
\EndFor
\end{algorithmic}
\end{algorithm}

To avoid loss of numerical precision due to round-off errors, it is advisable to perform the numerical summations after sorting the summands in ascending numerical order, thus adding elements from the smallest to the largest. For energy levels (bins) where no entries were observed in any on the intervals, the corresponding W array elements are always kept equal to zero.

The iterative method can also be used to solve the approximate self-consistent equations \ref{eq:poisson}, which converge significantly faster than the full ones. To this end, it suffices to modify the algorithm at lines 6, 12, and 26 of the pseudocode by removing the {\bf repeat} and {\bf until} statements, which makes variable $c$ unnecessary, and setting $r \gets 1$. The $W$ array obtained at convergence closely approximates the solution of the full set of self-consistent equations and is then used to initialize the latter for the final refinement.

\begin{figure}[h!]
    \centering
    \includegraphics[width=0.4\textwidth]{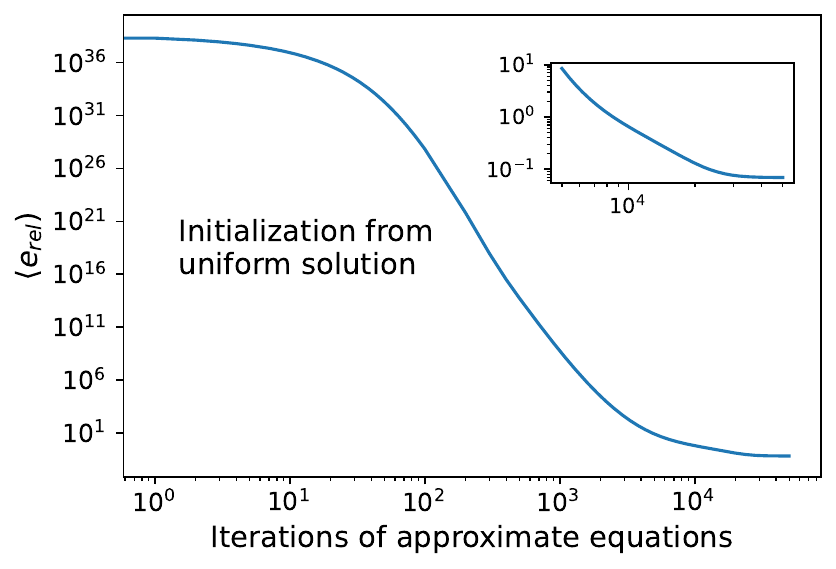}\\
    \includegraphics[width=0.4\textwidth]{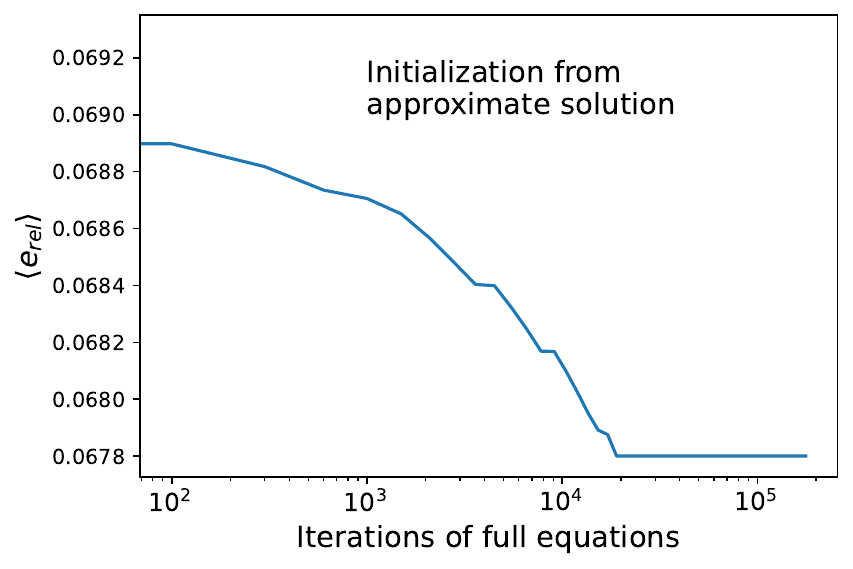}
    \caption{{\bf Convergence of the iterative procedure.} Average relative error of the two-step iterative solution of the self-consistent equations \ref{eqn:final}. (top) Starting from a uniform initial profile, the density of states of first iteratively evolved to solve the approximate equations \ref{eq:poisson}, which converge efficiently. (bottom) The converged solution of the first step is then refined by iteratively evolving it with the algorithm based on the full self-consistent equations \ref{eqn:final}.}
    \label{fig:convergence}
\end{figure}

Fig.~\ref{fig:convergence} illustrates the convergence of the two-step procedure for the $12\times12$ Ising model discussed in the main text Appendix A. For this application, we covered the admissible energy range with $\sim 150$ intervals of width $8$, corresponding to $m=3$ slack variables, and used a sampling depth of $1000$ for each histogrammed interval. The resulting density of states profile, $W(E)$, was compared with the one calculated exactly, $W^{\rm exact}(E)$. Considering the wide dynamic range of $W$ of the system, which spans $\sim 40$ orders of magnitude, we measured the convergence as the relative difference to the exact solution, averaged over all distinct energy levels: $\langle e_{\rm rel}\rangle = \langle | W(E) - W^{\rm exact}(E)| / W^{\rm exact}(E) \rangle_E$.

The top panel of Fig.~\ref{fig:convergence}  illustrates the convergence of the iterative algorithm based on the approximate equations \ref{eq:poisson}, initialized from a uniform (flat) $W$ profile. Because the exact profile spans $\sim 40$ orders of magnitude, the flat initial profile implies a very large initial error. In spite of this, the iterative procedure converges rapidly within few percents of $W^{\rm exact}(E)$. The converged solution of the approximate equations is then refined by using it to initialize the iterative solution of the full equations.

\section{Solving QUBO problems with parallel tempering methods}
\label{sec:Quantum_annealing}
Classical parallel tempering (PT) methods, also termed replica exchange methods (REMs) guarantee that the ground state manifold of QUBO Hamiltonians is sampled uniformly.
Thus, we resorted to PT/REMs as the ideal tools to carry out a stringent and transparent validation of our framework for reconstructing density-of-states.

The PT/REM-based optimization of a given QUBO Hamiltonian, ${\cal H}(\sigma_1, \dots ,\sigma_n)$, was set up as follows. The largest reduced temperature of the replicas, $k_B\, T_{\rm max}$, was picked by computing the root-mean-squared fluctuation of the QUBO energy for random spin configurations and then setting $k_B\, T_{\rm max}$ to be between 2 and 3 orders of magnitude larger than this characteristic energy fluctuation. The lowest reduced temperature, $k_B \, T_{\rm min}$, was picked by computing the root-mean-squared fluctuation of the QUBO energy for single-spin flips over random conformations and then setting  $k_B\, T_{\rm min}$ to be at least three orders of magnitude smaller than this characteristic single-spin energy fluctuation. The temperature range was then tentatively covered with $N_r$ replicas, where $N_r$ was customarily set equal to twice the square root of the number of degrees of freedom, $N_r=2 \sqrt{n}$ and spread across the entire range. Starting from this value, the number of replicas and their temperatures were iteratively refined with preliminary runs to ensure that the probability distributions of adjacent replicas intersected at approximately half their peak heights.

For the production runs, the elementary moves consisted of single spin flips applied to slack and non-slack spins. The two types of spin flips were alternated, i.e., after a non-slack spin flip, the next spin flip was restricted only to slack variables.
Ground state configurations were picked from the uncorrelated states sampled at the lowest-temperature replica. The autocorrelation time was computed from the decay of the time-lagged average overlap of the spin configurations.

\begin{figure*}[t!]
    \centering
    \includegraphics[width=0.9\linewidth]{./Figures/Supplementary_trial.pdf}
    \caption{{\bf Equilibrium linking probability of melts of self-assembled rings for $3$, $4$ and $5$ rings.} The central graph shows the linking probability, $P_{link}$, for a canonically-equilibrated ring melt filling a $5 \times 5 \times 4$ cubic lattice, considering only $3$ (blue line), $4$ rings (orange line) and $5$ rings (green line). The data points represent the average $P_{link}$ computed from $4$ reconstructions, and the shaded bands indicate the standard error of the mean. The callouts point to rendered sampled structures with the largest possible curvature (left column), average curvature (top, center), and minimal curvature (right column). The latter present a parity (even-odd) effect. The sampled configurations with 3 and 5 rings were free of concatenations. As illustrated, they feature stacked planar rings and a longer one that threads through them without concatenation, unlike configurations with 4 rings.}
    \label{fig_sup:linking_probability_34}
\end{figure*}

\section{Linking probability of melts with selected number of rings}

The non-monotonicity of the linking probability presented in the main text is a robust feature in that it does not depend on the fluctuating number of self-assembled rings, $n_{\rm rings}$. To establish this, we computed the linking probability for microstates with $n_{\rm rings}=3,\ 4$, and $5$ only. The $n_{\rm rings}$ selection was applied {\em a posteriori} on the states collected at the various sampled intervals.

The average linking probability for the equilibrated ring melt subject to the $n_{\rm rings}$ restriction is shown in Fig.~\ref{fig_sup:linking_probability_34}. As can be seen, the non-monotonicity is a robust feature because it is also present when considering a small number of rings.

\section{Average curvature vs bending rigidity}
\begin{figure}[!h]
    \centering
    \includegraphics[width=0.4\textwidth]{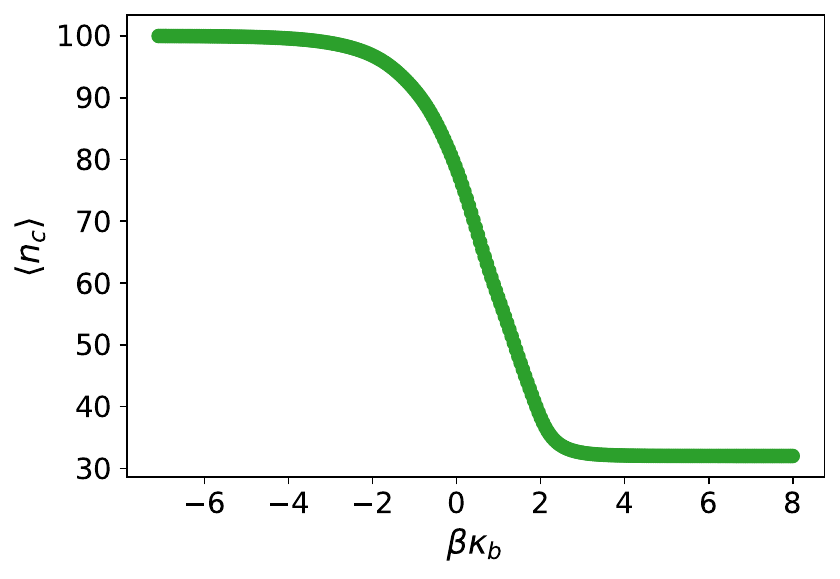}
    \caption{{\bf Average curvature of melts of self-assembled rings as a function of the reduced bending rigidity $\beta \kappa_b$.} Average curvature $\langle n_c \rangle$ for a canonically-equilibrated ring melt filling a $5 \times 5 \times 4$ cubic lattice. The data points represent $\langle n_c \rangle$ computed from $4$ reconstructions, and the shaded bands indicate the standard error of the mean. The considered range of $\beta \kappa_b$ spans from $-8$ to $+8$, i.e., from strongly favored to strongly suppressed corner turns.}
    \label{fig_sup:avg_number_corners}
\end{figure}

To complete the characterization of the canonical equilibrium properties of the ring melt, we used the density of states reconstruction and reweighting method to profile the average number of corner turns, $n_c$, versus the reduced bending rigidity, $\beta \kappa_b$. The resulting profile tying the two conjugate variables is shown in Fig.~\ref{fig_sup:avg_number_corners}.
